\documentclass[twocolumn,superscriptaddress,floatfix]{revtex4-2}

\usepackage{chemformula} 
\usepackage[T1]{fontenc} 
\usepackage{mathtools}
\usepackage{booktabs}
\usepackage{setspace}
\usepackage{hyperref}
\usepackage{physics}
\usepackage{mathtools}
\usepackage{amssymb}
\usepackage{url}
\DeclarePairedDelimiter\ceil{\lceil}{\rceil}

\hypersetup{
    colorlinks=true,
    citecolor=gray,
    filecolor=blue,
    linkcolor=blue,
    urlcolor=red
}

\begin{document}


\title[]
{Quantum computing within a bosonic context: Assessing finite basis effects on prototypical vibrational Hamiltonian spectra}

\author{Joachim Knapik}
\affiliation{Laboratoire Univers et Particules de Montpellier, Université de Montpellier, UMR-CNRS 5299, 34095 Montpellier Cedex, France}
\affiliation{ICGM, Univ Montpellier, CNRS, ENSCM, 34090 Montpellier, France}
\author{Bruno Senjean}
\affiliation{ICGM, Univ Montpellier, CNRS, ENSCM, 34090 Montpellier, France}
\author{Benjamin Lasorne}
\email{benjamin.lasorne@umontpellier.fr}
\affiliation{ICGM, Univ Montpellier, CNRS, ENSCM, 34090 Montpellier, France}
\author{Yohann Scribano}
\affiliation{Laboratoire Univers et Particules de Montpellier, Université de Montpellier, UMR-CNRS 5299, 34095 Montpellier Cedex, France}

\begin{abstract}

Quantum computing has recently been emerging in theoretical chemistry as a realistic avenue meant to offer computational speedup to challenging eigenproblems in the context of strongly-correlated molecular systems or extended materials. 
Most studies so far have been devoted to the quantum treatment of electronic structure, for which the transformation of fermionic operators into the qubit space is quite transparent.
In contrast, only a few were directed to the quantum treatment of vibrational structure, which at the moment remains not devoid of unknowns.
When simulating an anharmonic vibrational mode under bosonic second quantization, the biggest problem, which we analyze in detail in the present work, is the disruption of the resolution of the identity when truncating the infinite harmonic-oscillator basis set of primitive modals and how this alters the canonical commutation relation.
This may lead to serious incorrectness in the evaluation of the Hamiltonian matrix elements when assembling them from finite matrix products, which eventually entail a nonvariational behavior with respect to the finite size of the computational basis.
As we show here, a simple cure occurs to be obtained upon using Wick's normal order, for reasons that are not standardly evoked.
We also provide a detailed comparison between the boson-to-qubit unary and binary mappings under two different representations: one for the bosonic ladder operators within the harmonic-oscillator primitive basis, and one for the so-called $n$-mode representation within any type of computational basis.
In addition, we discuss the impact of choosing an adequate primitive basis set in terms of quantum computing with respect to its variational convergence efficiency (number of basis functions, hence of qubits) and as regards the magnitude of the 1-norm of the encoded Hamiltonian (a measure of the computational complexity of the quantum algorithm).
Such fundamental aspects are illustrated numerically on a
one-dimensional anharmonic Hamiltonian model corresponding to a symmetric double-well potential, of interest both for vibrational spectroscopy and chemical reactivity, and which is a challenging situation for numerical convergence due to fine tunneling splitting.

\emph{Keywords:} quantum computing; Hamiltonian mapping; qubit encoding; ladder operators and normal order; bosonic modes; vibrational structure; tunneling in spectroscopy and chemistry. 
\end{abstract}

\maketitle 

\section{Introduction}

Quantum-computing (QC) techniques are nowadays believed to be promising computational alternatives compared to classical ones as regards the simulation of molecular properties, especially when the size and complexity of the problem makes it intractable on a classical computer~\cite{doi:10.1021/acs.chemrev.8b00803}.
In particular, electronic-structure calculations have been known to already benefit from significant algorithmic progress in QC~\cite{PhysRevX.6.031007,RevModPhys.92.015003, lanyon_towards_2010,Papier,rossmannek_quantum_2023,PRXQuantum.2.020310,Romero_2019}.
In comparison, the simulations of molecular vibrational structure on a quantum computer are still in their infancy, especially for qubit-based devices.
Indeed, the study of vibronic spectra on bosonic devices started a decade ago~\cite{joshi2014estimating,
huh2015boson,
huh2017vibronic,
clements2018approximating,
hu2018simulation,
shen2018quantum,
chin2018quantum,
sparrow2018simulating,
wang2020efficient,
jahangiri2020quantum,
yalouz2021encoding,
wang2022hybrid,
dutta2024simulating,
malpathak2025simulating},
mainly focusing on the estimation of
Franck-Condon spectra using Gaussian boson sampling.
Simulations of infinite bosonic systems on finite qubit-based devices appear less natural~\cite{chiari2025abinitiopolaritonicchemistry}.
Hence, they are more recent and have been studied mostly because of the popularity of the latter~\cite{mcardle2019digital,
sawaya2019quantum,
Ollitrault_2020,
Sawaya_2020,
PhysRevA.104.062419,
nguyen2023description,
wang2023quantum,
bahrami2024particle,
olarte2024simulating,
majland2024vibrational,
trenev2025refining,
somasundaram2025quantum}.

When doing QC on a qubit-based digital quantum computer, one
often has to express the Hamiltonian within a second-quantized operator formulation via some kind of algebraic mapping that relies on the preliminary choice of a computational basis.
Such a formulation requires to know the action of the second-quantized operators with respect to the basis~\cite{Ollitrault_2020}.
As regards the electronic-structure problem, the so-called quantum-chemistry Hamiltonian has its universal second-quantized definition with standard one- and two-body integrals.
It relies in practice on the knowledge of a one-body basis set, the spin-orbitals, and the subsequent definition of creation and annihilation operators governing the occupation 
of each one-body spin-orbital involved within the many-body Slater determinants.
The later are then represented as occupation number vectors (ONVs), with entries having value zero if the spin-orbital is empty and one if occupied.
Hence, each spin-orbital in the ordered list is one-to-one mapped to a qubit.
Dealing with the many-fermion antisymmetric properties of Fermi-Dirac statistics with respect to permutations, according to the Pauli exclusion principle, is further ensured directly in the creation and annihilation operators
through the Jordan-Wigner mapping~\cite{jordan1928p}, which involves a nonlocal antisymmetrization ``pre-string'' of Pauli $Z$ matrices for the previous spin-orbitals within the ordered list with respect to the spin-orbital of interest.
This naturally turns the qubit string into an actual many-fermion ordered ONV that duly shares the antisymmetric properties of a Slater determinant represented as a normal-ordered product sequence of creation and annihilation operators acting on the fermionic vacuum state, according to Wick's theorem~\cite{Wick}.

In contrast, along the vibrational-structure side, such a one-to-one correspondence is less transparent
between a vibrational mode (degree of freedom), its modals (basic vectors), and a qubit-based mapping (operator encoding).
Different vibrational Hamiltonian representations can be used depending on the problem being addressed~\citep{PhysRevA.98.042312,PhysRevA.109.032612}.
A natural way to define a vibrational Hamiltonian within a bosonic perspective (a potentially-coupled system of assumedly spinless and distinguishable vibrational modes, each able in principle to accommodate an infinite number of bosonic quanta, \emph{i.e.}, mode-excitation quasiparticles: the vibrational bosons, sometimes referred to as ``vibrons'') is found upon using a relevant model and its standard primitive basis such as the harmonic-oscillator one.
Hence, the canonical commutation relation between the position and momentum operators yields a fully equivalent description in terms of standard bosonic ladder creation and annihilation operators. 
In addition, while this choice of basis is rarely the most compact one~\citep{Ollitrault_2020}, it is numerically convenient, since it allows one to get analytic integrals for evaluating all monomial functions of the position or momentum operators within a finite polynomial Taylor expansion.
Then, if one wants to use qubits here, the bosonic ladder operators should further be mapped to conventional strings of two-level raising and lowering Pauli operators, and various types of such algebraic or numeral boson-to-qubit encodings (in particular, unary or binary) have been developed for this purpose~\citep{SommaIJQI,Sawaya_2020,huang2022qubitization}, which are recalled in the present work.

Finally, an important distinction should be made between the electronic- and vibrational-structure problems as regards second quantization.
While the number of fermionic modes (spin-orbitals) of the former should in principle be infinite, but
each mode is exactly empty or occuppied (either 0 or 1 physical particle),
the number of bosonic modes of the latter (molecular vibrations) is finite (fixed by the interatomic graph connection of the molecular system), while the harmonic-oscillator basis has an infinite number of elements (modals) labelled by the number of vibrational bosons within the mode (0, 1, 2, ..., up to an  infinite number of excitation quasiparticles).

Of course, in practical simulations (either classical or quantum) one
always has to truncate the Hilbert primitive basis.
Increasing its size in the electronic case typically leads to a variational convergence behavior with a smooth decrease of the average energy from above to a lower bound.
However, truncating the primitive basis can potentially be much more problematic in the vibrational case within a second-quantized context.
Indeed, special care has to be taken as regards the ordering of products of the finite matrix representations of the bosonic ladder operators involved within the ``polyladder'' vibrational Hamiltonian matrix expansion. 
If not, even in the harmonic case, one thus simulates a fictitious system that is not only a finite variational approximation of the linear harmonic-oscillator model but -- in fact -- something else, with spurious eigenstates and eigenvalues, specifically named as the ``truncated harmonic oscillator'' model by Buchdahl in the 1960s, which actually implies a very modification of the canonical commutation rules~\citep{10.1119/1.1974004}.
Prototypical upper-bounded bosonic-like models such as the vibrational Morse oscillator, which really has a finite number of bound eigenstates, have been rationalized adequately over many decades along similar algebraic lines~\citep{CARVAJAL2000105}.
Yet, the numerical effect of such a predicament when truncating an anharmonic oscillator model that is not upper-bounded, for example a quartic double well~\cite{mitoli2025anharmonic} as illustrated in the present work, is not easily predictable and truly deserves some detailed analysis.

At this stage, it remains evident that we may still encounter unexpected formal or numerical traps when applying QC to vibrational-structure problems without care, which justifies that some investigations of basic one-dimensional models must be carried out dutifully before going further.
Along this line, various types of prototypical one-dimensional vibrational models have been explored recently within the QC framework~\citep{PhysRevResearch.6.013032}.
Herein, we focus on a simple but rich one-dimensional vibrational model
with a double-well potential exhibiting a small splitting of its eigenvalues due to fine tunneling (wavenumber difference lower than one reciprocal centimeter), of relevance for vibrational spectroscopy (\emph{e.g.}, ammonia inversion~\citep{Dennison-Uhlenbeck_pr1932,RICARDOLETELIER1997247,Nguyen_jpcl2024}) or chemical reactivity (\emph{e.g.}, hydrogen transfer in malonaldehyde~\citep{lauvergnat2023malonaldehyde}). Such a model has a highly nontrivial convergence behavior as regards its lowest eigenvalues, which makes it perfectly adequate to highlight the aforementioned consequences of the basis truncation.

The present work is expected to benefit as a preliminary \emph{vade-mecum} to researchers already doing vibrational structure with classical computing but aiming at exploring the capabilities offered by QC, as well as to researchers already doing QC for electronic structure and interested in generalizing their algorithms to vibrational structure.
It is outlined as follows.
First, some detailed conceptual background is given in the next section.
Especially, we provide a detailed comparison between the boson-to-qubit unary and binary mappings under two different representations: one for the bosonic ladder operator within the harmonic-oscillator primitive basis, and one for the so-called $n$-mode representation within any type of computational basis.
We then provide essential information about the model and relevant computational details.
The section before last is dedicated to results and discussion that highlight what can be taken as good practice
(so as to avoid unexpected traps) for vibrational simulations when using QC, while at the same time assessing the question of the choice of a good basis, which can play a significant role on the efficiency of quantum algorithms, for instance by modifying the 1-norm of the Hamiltonian to be simulated.
We finish with concluding remarks and outlook.
Extra information is given in Appendices for completion.

\section{The vibrational problem on a  quantum computer}

In the present section, we address the mappings of different representations of a vibrational problem to be solved by qubit-based quantum computers, with some particular focus on the binary mapping.
Additional details are provided in Appendix~\ref{app:ladder}.
We shall show that a second-quantized form of the vibrational Hamiltonian in terms of bosonic ladder operators proves essential for further encoding its action with respect to a primitive basis set but implies some precautions when the basis is truncated within a computational context.

\subsection{Direct and compact mapping of bosonic ladder operators}
\label{sec:mappings}

The generic vibrational problem expressed in terms of $L \geq 1$ normal modes corresponds to the following multidimensional first-quantized  Hamiltonian in position representation (using mass-weighted rectilinear normal coordinates, and assuming $\hbar=1$), 
\begin{eqnarray}\label{eq:ham}
\hat{H} = - \dfrac{1}{2}\sum_{l=1}^L \dfrac{\partial^2}{\partial x_l^2} + V(x_1, \hdots, x_L) \quad.
\end{eqnarray}
The most general second-quantized formulation maps each pair of position, $\hat{x}_l$, and momentum,
$\hat{p}_l$, self-adjoint operators to an equivalent -- non-self-adjoint, but mutually adjoint -- pair of bosonic operators, $\hat{b}_l$ and $\hat{b}^\dagger_l$, via
\begin{eqnarray}
\hat{x}_l &=& \dfrac{1}{\sqrt{2}}(\hat{b}_l + \hat{b}^\dagger_l) \quad, \nonumber \\
\hat{p}_l &=& \dfrac{1}{i\sqrt{2}}(\hat{b}_l - \hat{b}^\dagger_l) \quad,
\label{eq:canorel}
\end{eqnarray}
such that the canonical commutation relation,
\begin{equation}
[\hat{x}_l, \hat{p}_k] = i\hat{1} \delta_{lk} \quad,
\end{equation}
translates into a standard bosonic commutation relation, 
\begin{equation}
[\hat{b}_l, \hat{b}^{\dagger}_k] = \hat{1} \delta_{lk} \quad,
\label{eq:bosocom}
\end{equation}
where $\delta_{lk}$ denotes the Kronecker symbol.
In the multidimensional direct-product harmonic-oscillator basis, these bosonic operators occur to be the second-quantized harmonic ladder -- creation and annihilation -- operators that act, for each degree of freedom, $l=1, \hdots, L$, as
\begin{eqnarray}\label{eq:ladder_op}
\hat{b}^\dagger_l \ket{n_1\hdots n_l \hdots n_L} &=& \sqrt{n_l +1} \ket{n_1\hdots n_l+1 \hdots n_L} \quad, \nonumber\\
\hat{b}_l \ket{n_1\hdots n_l \hdots n_L} &=& \sqrt{n_l} \ket{n_1\hdots n_l-1 \hdots n_L} \quad,
\end{eqnarray}
where $\lbrace \ket{n_1\hdots n_l \hdots,n_L} \rbrace$ is the set of the $L$-sized occupation number vectors (ONVs) for the $L$ normal modes.
In contrast to the electronic-structure problem where the fermionic or hard-bosonic occupations can only be 0 or 1, 
the bosonic occupation numbers, $n_l$, can range from zero to infinity in principle and -- in practice -- be as large as the finite size of the basis, futher denoted by $M_l$ for the $l$-th mode.
Hence, a greater-than-one bosonic occupation number cannot be encoded directly within a single qubit, as generally done
in quantum chemistry using the Jordan--Wigner mapping for fermions~\cite{jordan1928p}.
While the simulation of a vibrational problem on a physical bosonic quantum device thus appears more appealing~\cite{joshi2014estimating,
huh2015boson,
huh2017vibronic,
clements2018approximating,
hu2018simulation,
shen2018quantum,
chin2018quantum,
sparrow2018simulating,
wang2020efficient,
jahangiri2020quantum,
yalouz2021encoding,
wang2022hybrid,
dutta2024simulating,
malpathak2025simulating}, qubits are still used more commonly, and we rather consider for the moment $L$ registers of several qubits as our focus herein to encode the greater-than-one occupation numbers of the $L$ bosonic modes for digital quantum computations. 

Below, we shall further restrict ourselves to one-dimensional models, such that $L=1$, which fully deserve preliminary attention on their own right.

\subsubsection{Direct mapping}

The simplest multi-qubit-based mapping for bosons~\cite{SommaIJQI} is called the direct or unary mapping~\cite{RevModPhys.92.015003,Sawaya_2020,Ollitrault_2020}
-- also referred to as one-to-one~\citep{{huang2022qubitization}}, one-hot~\citep{Chancellor_2019,Sawaya_2020}, or single-excitation subspace~\citep{PhysRevA.91.062309} --,
which consists in encoding the occupation of the $l$-th mode onto $M_l$ qubits.
Interestingly enough, such a unary-mapping approach can be traced back at least to 1997, during the pre-QC-era, with the seminal work of Thoss and Stock about the second-quantized theoretical treatment of vibronic phenomena~\citep{PhysRevLett.78.578,PhysRevA.59.64}, which -- together with the spin-$S$ mapping of finite $(2S+1)$-state subspaces, earlier explored by Meyer and Miller~\citep{10.1063/1.438598} in 1979 -- gave rise to what is now known as the MMST mapping within its own scientific context: namely, nonadiabatic semiclassical dynamics based on spin-boson diabatic Hamiltonian models.
A detailed and illuminating review of algebraic dictionaries among various languages relating prototypical systems and models of different types within a QC context has been proposed by Batista and Ortiz in 2004~\citep{Batista01012004}.
Within the unary mapping, all qubit-states are set to 0 but the state of the $n_l$-th qubit, which is set to 1.
This mapping is particularly noneconomic in terms of number of qubits, as one
requires $\sum_l M_l$ qubits in total and a very sparse exploration of the symbolic Fock space: only a linear number of configurations from the exponentially-many ones is physically relevant.
However, it is quite transparent as regards the transcription of the action of the multi-level bosonic ladder operators in terms of strings of two-level qubit operators.
Indeed,  for any mode $l$, the restricted bosonic creation operator (projection of $\hat{b}_l^\dagger$ onto the finite harmonic basis of size $M_l$),
\begin{eqnarray}\label{eq:op_ladder}
\hat{b}_l^\dagger \rightarrow \hat{b}_{M_l}^{\dagger} =
 \sum_{r_l=0}^{M_l-2} \sqrt{r_l + 1} \, \ket{r_l+1}\bra{r_l}\quad ,
\end{eqnarray}
is isomorphic, within the unary mapping, to Kronecker direct products of raising and lowering qubit operators describing occupation transitions between first-neighbor pairs -- also known as Ising interactions~\citep{SommaIJQI} -- 
\begin{eqnarray}\label{eq:op_unary}
 \hat{b}_{M_l}^{\dagger} = \sum_{r_l=1}^{M_l-1} \sqrt{r_l} \, \hat\sigma^-_{r_l+1}  \otimes  \hat\sigma^+_{r_l} \quad,
\end{eqnarray}
where tensor products of identity matrices have been omitted for the sake of compactness,
and we label the qubits starting from 1 in this work.
This expansion can be regrouped into only $O(M_l)$ local Pauli strings upon using the following algebraic equivalences,
\begin{eqnarray}\label{eq:projectors_to_pauli}
\hat\sigma^- &\equiv& \ket{1}\bra{0} \equiv \dfrac{1}{2}(X - iY) \quad, \nonumber \\
\hat\sigma^+ &\equiv& \ket{0}\bra{1} \equiv \dfrac{1}{2}(X + iY) \quad, \nonumber \\
I^- &\equiv& \ket{1}\bra{1} \equiv \dfrac{1}{2}(I - Z) \quad, \nonumber \\
I^+ &\equiv& \ket{0}\bra{0} \equiv \dfrac{1}{2}(I + Z) \quad,
\end{eqnarray}
where $X$, $Y$, and $Z$ notations are the standard matrix representations of the $\hat\sigma_x$, $\hat\sigma_y$, and $\hat{\sigma}_z$ Pauli operators (also known as Pauli gates within a QC context).
This aspect is further detailed in Appendix~\ref{app:Hqubit}.
Note that Pauli operators and their standard matrix representations can be identified for notational simplicity when there is no ambiguity.

\subsubsection{Compact mapping}

A more qubit-economic mapping is provided by the so-called compact mapping~\cite{RevModPhys.92.015003,Sawaya_2020}, which is equivalent to a numeral binary encoding of each integer number $n_l$. 
Such an encoding has also been used in the context of electronic-structure theory to simulate a block of fixed-particle number in the second-quantized Hamiltonian, known as the configuration-interaction matrix~\cite{toloui2013quantumalgorithmsquantumchemistry,babbush2017exponentially,shee2022qubit},
or to simulate non-interacting systems such as in quantum density-functional theory~\cite{senjean2023toward}, H\"uckel theory~\cite{singh2024huckel}, or the free-fermion problem~\cite{stroeks2024solvingfreefermionproblems}.
There, only $K = \ceil{\log_2(M_l)}$ qubits, where $\ceil{\cdot}$ is the ceiling function, are required for the mode $l$ but at the expense of
having $O(M_l \log_2(M_l))$ nonlocal Pauli strings when decomposing the ladder operators
in Eq.~(\ref{eq:op_ladder})
with
\begin{eqnarray}\label{eq:binary_rep}
\ket{r_l} =  \ket{\eta_{K}} \ket{\eta_{K-1}} \hdots \ket{\eta_1} \quad,
\end{eqnarray}
and the binary representation 
$r_l = \eta_{K} 2^{K-1} + \eta_{K-1} 2^{K-2} + \hdots + \eta_{1} 2^{0}$ where $\eta_i \in \{0,1\}$ for $1 \leq i \leq K$ (as in the previous section, we label the qubits starting from 1).
In practice, this implies to choose some large enough value of $K$ (number of qubits) such that $M_l=2^K$ will be the actual size of the primitive basis for mode $l$.
Combining Eqs.~(\ref{eq:projectors_to_pauli}-\ref{eq:binary_rep}), each first-neighbor transition (jump) operator within the sum in Eq.~(\ref{eq:op_ladder}) can be decomposed into $O(2^K) = O(M_l)$
nonlocal Pauli strings, thus leading to $O(M_l^2)$ Pauli strings for the ladder operator under compact encoding, as mentioned by McArdle and coworkers.\cite{mcardle2019digital} 
Here, we found that this number can
actually be reduced to $O(M_l\log_2(M_l))$ only, due to the presence of $(2^{K} - (K+1) )2^{K} = M_l^2 - (\log_2(M_l)+1)M_l$ redundant Pauli strings, as shown in Appendix~\ref{app:Hqubit}.

In comparison with the direct encoding, the binary one leads to more complicated expressions.
Indeed, the decomposition of the restricted bosonic creation operator $\hat{b}_l^\dagger \rightarrow \hat{b}_{M_l}^{\dagger}$ into a linear combination of Pauli string is obtained from a recursive relation, where
we use, for a system represented with $K$ qubits, the shorthand notation $\hat{b}_{K}^{\dagger} \equiv \hat{b}_{2^K}^\dagger = \hat{b}_{M_l}^\dagger$ to be consistent with Ref.~\citep{huang2022qubitization}
but with a different ordering, as in Eq.~(\ref{eq:binary_rep}),
\begin{eqnarray}
        \hat{b}^{\dagger}_{K} &=& \sum\limits_{i=1}^{2^{K-1}-1}  \hat{c}_{i}^{(1,K-1)} \otimes \left(
        \sqrt{i} \, {I}^{+}_{K}  
        +
        \sqrt{2^{K-1} + i} \, {I}^{-}_{K}\right) \nonumber \\
        && +
        \sqrt{2^{K-1}} \, \hat\sigma^+_{1}\otimes\hat\sigma^+_{2}\otimes \hdots \otimes \hat\sigma^-_{K} \quad.
 \label{eq:Binary_bdag}
\end{eqnarray}
In this construction, the operators $\hat{c}_{i}^{(1,K-1)}$ are to be understood as those entering the expression of the restricted bosonic creation operator of a system now represented with $K-1$ qubits in the following manner,
\begin{eqnarray}
        \hat{b}^{\dagger}_{K-1}
        = \sum_{i=1}^{2^{K-1}-1} \sqrt{i} \, \hat{c}_i^{(1,K-1)}\quad .
        \label{eq:Binary_bdag_minus}
\end{eqnarray}
In practice, $\hat{b}^{\dagger}_{1} = \hat\sigma^-_{1}$, which induces
\begin{equation}
    \hat{b}^{\dagger}_{2} = \hat\sigma^-_{1}
   \otimes   {I}^{+}_{2} + \sqrt{2}\, \hat\sigma^+_{1}  \otimes \hat\sigma^-_{2} + \sqrt{3}\, \hat\sigma^-_{1} \otimes  {I}^{-}_{2}  \quad, \label{eq:Ex_binary_bdag2}
\end{equation}
and then,
\begin{eqnarray}
    \hat{b}^{\dagger}_{3}   & =  &  \hat\sigma^-_{1} \otimes {I}^{+}_{2} \otimes {I}^{+}_{3} + \sqrt{2} \, \hat\sigma^+_{1}\otimes \hat\sigma^-_{2}  \otimes  {I}^{+}_{3} + \sqrt{3} \, \hat\sigma^-_{1} \otimes {I}^{-}_{2} \otimes {I}^{+}_{3} \nonumber \\
    && + 
    \sqrt{4}\, \hat\sigma^+_{1} \otimes \hat\sigma^+_{2}  \otimes  \hat\sigma^-_{3} 
     + \sqrt{5}\, \hat\sigma^-_{1} \otimes {I}^{+}_{2} \otimes {I}^{-}_{3}  \nonumber \\
     & & + \sqrt{6}\,  \hat\sigma^+_{1} \otimes \hat\sigma^-_{2}  \otimes {I}^{-}_{3}  + \sqrt{7}\,\hat\sigma^-_{1} \otimes {I}^{-}_{2} \otimes   {I}^{-}_{3} \quad.
\label{eq:Ex_binary_bdag3}
\end{eqnarray}
A concrete example
explaining the origin of Eq.~(\ref{eq:Binary_bdag}) is provided in Appendix~\ref{app:origin_binary}.
Then, the square-root factors are regrouped to produce the actual coefficients of the Pauli strings, $\lambda_{i}$, which are implied in the expansion $\hat{H} = \sum\limits_{i}\lambda_{i} \hat{P}_{i}$, where $ \hat{P}_{i}= \bigotimes\limits_{j=1}^K \hat\sigma_{j} $ and $\hat\sigma_{j} \in \{ {I}, {X}, {Y}, {Z} \}$.

\subsection{Mapping occupation number vectors to computational bases}

Let us now turn to another representation of the vibrational problem.
The first-quantized Hamiltonian in Eq.~(\ref{eq:ham})
can be projected to any orthonormal and supposedly complete Hartree-product basis of the many-body Hilbert space,
$\lbrace \bigotimes_{l=1}^L  \ket{\varphi_{k_l}} \rbrace$;
not only the harmonic-oscillator primitive basis, but potentially a variationally contracted one of VSCF (vibrational self-consistent field) type, aimed at being more compact.
Such a formulation has the advantage of making us able to rewrite the Hamiltonian
in Eq.~(\ref{eq:ham}) under a hierarchical many-body cluster-expansion form (also known as the $n$-mode representation) that reads
\begin{eqnarray}\label{eq:ham_olli}
\hat{H} &=& \sum_{l=1}^L \sum_{k_l,h_l}^{M_l}
\bra{\varphi_{k_l}}\hat T_l + \hat V^{(1)}_{l}\ket{\varphi_{h_l}} \ket{\varphi_{k_l}}\bra{\varphi_{h_l}} \nonumber \\
&&+
 \sum_{l<m}^L \sum_{k_l,h_l}^{M_l}  \sum_{k_m,h_m}^{M_m} 
\bra{\varphi_{k_l}\varphi_{k_m}}\hat V^{(2)}_{l,m}\ket{\varphi_{h_l}\varphi_{h_m}}
\nonumber \\
&&\times 
\ket{\varphi_{k_l}\varphi_{k_m}}\bra{\varphi_{h_l}\varphi_{h_m}}
+
\hdots \quad,
\end{eqnarray}
where the first term is the one-body problem, which will be the main focus of the present work (where $L=1$ herein because we shall be considering a one-dimensional model system). 
In the above equation, $\varphi_{k_l}$ can be a one-body primitive (\emph{e.g.}, harmonic-oscillator) basis function or a variationally optimized modal wavefunction obtained within a VSCF scheme.

Interestingly, this form can also be recast using symbolic
creation and annihilation operators, $\hat{a}^\dagger_{k_l}$ and $\hat{a}_{h_l}$, defined such that $\hat{a}^\dagger_{k_l}\hat{a}_{h_l} \equiv \ket{\varphi_{k_l}}\bra{\varphi_{h_l}}$, according to the Jordan map, 
\begin{eqnarray}\label{eq:ham_olli}
\hat{H} &=& \sum_{l=1}^L \sum_{k_l,h_l}^{M_l}
\bra{\varphi_{k_l}}\hat T_l + \hat V^{(1)}_{l}\ket{\varphi_{h_l}}\hat{a}^\dagger_{k_l}\hat{a}_{h_l} \nonumber \\
&&+
 \sum_{l<m}^L \sum_{k_l,h_l}^{M_l}  \sum_{k_m,h_m}^{M_m} 
\bra{\varphi_{k_l}\varphi_{k_m}}\hat V^{(2)}_{l,m}\ket{\varphi_{h_l}\varphi_{h_m}}\hat{a}^\dagger_{k_l}\hat{a}^\dagger_{k_m}\hat{a}_{h_l}\hat{a}_{h_m} \nonumber \\
&&
+
\hdots \quad,
\end{eqnarray}
which
is the analogue of the second-quantized electronic-structure problem~\cite{Ollitrault_2020,trenev2025refining} (appart from the absence of nonlocal Jordan-Wigner ``pre-string'' of Pauli $Z$ within the detailed definition of the creation and annihiliation operators, together with the general assumption of distinguishability  of the vibrational modes corresponding to different one-body bases).

As noticed by Ollitrault {\it et al.}~\cite{Ollitrault_2020} and Trenev {\it et al.}~\cite{trenev2025refining}, when considering the direct/unary mapping of the ONVs,
the operators $\hat{a}^\dagger_{k_l}(\hat{a}_{k_l})$ correspond to the
standard qubit mode-(de)excitation operators, also called hard-boson creation (annihilation) operators -- which identify to the local spin-$1/2$ Pauli lowering (raising) operators on ``site'' $k_l$, \textit{i.e.}, $\hat{a}^\dagger_{k_l} \equiv \hat\sigma^-_{k_l}$ and $\hat{a}_{k_l} \equiv \hat\sigma^+_{k_l}$, within the symbolic ``lattice''.
They are thus fully related to the direct/unary mapping introduced previously for the harmonic bosonic ladder operators in Eqs.~(\ref{eq:ladder_op}-\ref{eq:op_unary}) but with respect to a potentially more general basis,
such that
\begin{eqnarray}
\hat{a}^\dagger_{k_l}\hat{a}_{h_l} \ket{\varphi_{j_l}} &=& \hat{a}^\dagger_{k_l}\hat{a}_{h_l} \ket{0_{M_l} \hdots 1_{j_l} \hdots 0_{1}} \nonumber \\
& =&
\hat{a}^\dagger_{k_l} \delta_{h_l j_l} \ket{0_{M_l} \hdots 0_{1}} \nonumber \\
& =& 
\delta_{h_l j_l} \ket{0_{M_l} \hdots 1_{k_l} \hdots 0_{1}} \nonumber \\
&=& \delta_{h_l j_l}\ket{\varphi_{k_l}}.
\end{eqnarray}

Now, let us return to the compact/binary mapping. Expressing the transition operator $\ket{\varphi_{k_l}}\bra{\varphi_{h_l}}$ in terms of
elementary raising and lowering operators is highly non-trivial in comparison.
Indeed, such an expression has been derived in the context of quantum density functional theory~\cite{senjean2023toward} and reads
\begin{eqnarray}
| \varphi_{k_l} \rangle \langle \varphi_{h_l} |& =& \prod_{i = 1}^{K}   \left(\hat{\sigma}^+_{i} \hat{\sigma}^-_{i}\right)^{(1 - \eta_{i}^{k_l})(1 - \eta_{i}^{h_l}) } \left( \hat{\sigma}_{i}^+\right)^{(1 - \eta_{i}^{k_l})\eta_{i}^{h_l}}
\nonumber \\
&& \times 
\left(\hat{\sigma}_{i}^{-}\right)^{\eta_{i}^{k_l}(1 - \eta_{i}^{h_l}) } \left(\hat{\sigma}_{i}^{-}\hat{\sigma}_{i}^+ \right)^{\eta_{i}^{k_l} \eta_{i}^{h_l}},
\label{eq:binary_phik_phil}
\end{eqnarray}
where we used the binary representation of the ONVs [see Eq.~(\ref{eq:binary_rep})], $\ket{\varphi_{k_l}} = \vert \eta_{K}^{k_l}\rangle \hdots \vert \eta_1^{k_l}\rangle$ with $\eta_i^{k_l} \in \lbrace 0, 1 \rbrace$.
As far as we know, no general relation between the operators $\hat{a}^\dagger_{k_l}$ and $\hat{a}_{k_l}$, taken separately, and the
elementary raising and lowering operators has been derived yet within the binary encoding.
Now, when using the harmonic-oscillator primitive basis, we know that the transition operators expressed in terms of the bosonic ladder operators satisfy, per construction,  
\begin{eqnarray}
\label{eq:dyadharm}
\ket{\varphi_{k_l}}\bra{\varphi_{h_l}} = \dfrac{(\hat{b}^\dagger)^{k_l}}{\sqrt{k_l !}} \ket{0} \bra{0} \dfrac{(\hat{b})^{h_l}}{\sqrt{h_l !}}  \quad .
\end{eqnarray}
Hence, for $K$ qubits and within the binary encoding [see Eq.~(\ref{eq:Binary_bdag})], the corresponding raising and lowering operators should be identified to
\begin{eqnarray}
\hat{a}^\dagger_{k_l} &\equiv& \dfrac{(\hat{b}_K^\dagger)^{k_l}}{\sqrt{k_l !}} \ket{0}\bra{0}\quad ,\nonumber \\
\hat{a}_{k_l}  &\equiv& \ket{0}\bra{0}
\dfrac{(\hat{b}_K)^{k_l}}{\sqrt{k_l !}} \quad ,
\end{eqnarray}
where $\ket{0}$ represents the harmonic state zero (bosonic vacuum: all $K$ qubits are in state $\ket{0}$) and
the division by $\sqrt{k_l !}$ is here to cancel the factors that arise when applying the harmonic ladder operator to a given state [see Eq.~(\ref{eq:ladder_op})].
This finally leads to the following expression within the binary encoding,
\begin{eqnarray}
| \varphi_{k_l} \rangle \langle \varphi_{h_l} | 
\equiv \hat{a}^\dagger_{k_l} \hat{a}_{h_l}
\equiv \dfrac{(\hat{b}_K^\dagger)^{k_l}}{\sqrt{k_l !}} \ket{0}\bra{0}\dfrac{(\hat{b}_K)^{h_l}}{\sqrt{h_l !}}\quad,
\end{eqnarray}
which may seem more complicated than Eq.~(\ref{eq:binary_phik_phil}) but does not require prior knowledge about the binary representation of the ONVs (\emph{i.e.}, the values of the numbers $\lbrace \eta_i \rbrace$).
Alternatively, one could use operators that are analogous to the bosonic ladder operators but that do not rely on the harmonic-oscillator primitive basis, \emph{i.e.},
\begin{eqnarray}
        \hat{d}^{\dagger}_{K} &=& \sum\limits_{i=1}^{2^{K-1}-1}  \hat{c}_{i}^{(1,K-1)} \otimes \left(
        {I}^{+}_{K}  +
        {I}^{-}_{K}\right) +
        \hat\sigma^+_{1}\otimes\hat\sigma^+_{2}\otimes \hdots \otimes \hat\sigma^-_{K}, \nonumber \\
 \label{eq:Binary_ddag}
\end{eqnarray}
which is the same as Eq.~(\ref{eq:Binary_bdag}) but without the square-root factors.
Finally, using Eq.~(\ref{eq:projectors_to_pauli}), we thus obtain a final expression for $\hat{a}^\dagger_{k_l} $ ($\hat{a}_{k_l}$) in terms of elementary raising and lowering operators,
\begin{eqnarray}
\hat{a}^\dagger_{k_l} &\equiv& (\hat{d}_K^\dagger)^{k_l} \times \bigotimes\limits_{i=1}^{K} I_i^+ \quad .
\end{eqnarray}
Again, such an expression is obviously much more complicated than within the unary encoding.

Now, and much as for the electronic-structure problem, we are facing the potential difficulty of evaluating numerically the integral factors in front of the qubit operators in Eq.~(\ref{eq:ham_olli}), within the cluster-expansion hierarchy of one-body terms, two-body-terms, \emph{etc.}
There is a plethora of approaches for that, for example analytic quadrature techniques (according to specific primitive basis types). 
Numerical integral evaluations on grids are another option but they become rapidly prohibitive as the number of degrees of freedom increases, which is a striking example of what is known as the ``curse of dimensionality'' in vibrational-structure theory. 
Also, in contrast with electronic-structure theory -- where the expression of the potential energy is known analytically from first principles (Coulombic interactions between point charges) -- it must be stressed here that, in vibrational-structure theory, the global potential-energy function is only known first as a mere ``numerical object'' that is never given in advance and requires the evaluation of a large enough sample of its numerical values on a grid -- obtained from numerous Born-Oppenheimer electronic-structure computations, or even beyond -- to be made preliminarily and further fitted to a functional parametric ansatz (for example an anharmonic model). Such a requirement often represents some tedious work to be carried out first as a pre-processing step.

Within the present study, we shall rather consider from the onset a one-dimensional potential-energy model involving polynomial powers of a single coordinate,  up to degree four.
Using a description in terms of the harmonic ladder operators $\hat{b}$ and $\hat{b}^{\dagger}$ [see Appendix~\ref{app:ladder}] will provide analytic expressions of the aforementioned integrals. 
At this stage it is perhaps important to make the following remark: while they are not totally disconnected from the overall mapping procedure, there are two very different types of second quantization here and they should not be confused together.
The physical bosonic one with its harmonic ladder operators ($\hat{b}$ and $\hat{b}^{\dagger}$) is there for two reasons: it is formally a natural algebraic way of expressing the vibrational Hamiltonian under a bosonic form when using a harmonic-oscillator primitive basis; 
it also provides an easy avenue for numerically evaluating the matrix elements of the potential energy when expressed as a polynomial expansion with respect to the normal coordinate(s).
Then, the second-quantized operators within the $n$-mode representation ($\hat{a}$ and $\hat{a}^\dagger$)
allows us to navigate through any type of computational basis.
In both cases, when turning to the encoding of these operators onto a qubit-based device,
they can be expressed as a linear combination of
hard-bosonic ones ($\hat{\sigma}^+$ and $\hat{\sigma}^{-}$) under different manners, depending on the use of a unary or binary encoding.

\subsection{Dealing with a finite basis}

Below, we shall discuss the two types of errors that arise when truncating the primitive basis set within a QC context for vibrational Hamiltonian models with \emph{a priori} unknown eigensolutions.

First, there is a well-known variational error that comes from using a finite computational basis to describe an intrinsically infinite Hilbert space under first quantization.
The Hamiltonian operator is thus represented in practice by a projected matrix of finite size.
This is unavoidable and manifests itself both in classical computing and QC, for the vibrational- or the electronic-structure eigenproblem.
Such an error is tamed by the Rayleigh-Ritz variational principle: as regards the energy of the ground eigenstate, the error is supposed to decrease smoothly and monotonically from above to a lower bound when the size of the primitive basis increases, thus always yielding well-behaved convergence properties in principle but at some potentially significant computational cost.
In this context, choosing a ``good primitive basis set'' for the problem at stake, \emph{i.e.}, a parametrization that ensures efficient variational convergence properties as concerns the numerics, is an essential aspect;
this will be addressed and illustrated with examples discussed in Section~\ref{sec:resdisc}.

Second, there is a second source of error, also due to the effect of basis truncation, but that is not variational.
It is not so well-known within the QC community, since it happens only for the vibrational-structure problem due to the intrinsically infinite nature of the bosonic Fock space under second quantization.
Such a formal error can be interpreted as an incomplete resolution of the identity when assembling products of finite intermediate matrices involved in the final series expansion of the finite Hamiltonian matrix.

While it has been shown for a while that a finite-matrix representation based on Wick's normal-ordered products of ladder operators is the required prescription within a harmonic context [see Appendix~\ref{app:ladder}] -- although this is not the original incentive for it -- the numerical effect of this error is not trivial for a Hamiltonian that is beyond harmonic, and this remains of interest to be explored from a more general perspective.
Indeed, one may be tempted to assume that such an error is marginal and will have a negligible effect on the final eigenvalues.
However, we shall show that it actually can break variationality for the ground state quite dramatically and within an unpredictable manner.
Hence, duly re-expressing the Hamiltonian second-quantized operator expansion according to Wick's normal order, which requires some additional pre-processing work, is a preliminary step that should never be circumvented; this will be illustrated numerically in Section~\ref{sec:resdisc}.

\section{Model and computational details}

In order to assess the aforementioned issues against a practical situation of spectroscopic and chemical interest, we shall consider a prototypical one-mode model with a symmetrical double well describing fine tunneling splitting~\cite{mitoli2025anharmonic}.
Its nondimensionalized Hamiltonian reads
\begin{equation}
\label{eq:hleft}
    \hat{H} = \hat{h}_{0} - \frac{1}{2\ell}\hat{x}^{3} + \frac{1}{8\ell^{2}}\hat{x}^{4} \quad, \hspace{1cm} \hat{h}_{0} = \frac{1}{2}(\hat{p}^{2}+\hat{x}^{2}) \quad,
\end{equation}
where $\hat{h}_{0}$ is the harmonic approximation of $\hat{H}$ around the bottom of the left well (the origin: $x=0$).
The (dimensionless length) width of the barrier is $\delta = 2 \ell$ and its (dimensionless energy) height is $ \eta = \frac{\ell^{2}}{8}$.
Here, we chose $\ell=4$~\cite{Burghardt_2022}.
Note that some energy rescaling will be considered hereafter for making things more concrete, with a harmonic frequency $\omega \equiv 0.0091127$ hartree $\equiv 2000$ $\text{cm}^{-1}$, which yields a barrier height of $4000$ $\text{cm}^{-1}$.

For emulating a QC situation, the qubit-operator representation of the Hamiltonian was generated using $\textsc{Qiskit}$~\citep{qiskit2024}, with a binary encoding of the creation and annihilation operators~\cite{huang2022qubitization} [further exemplified in Appendix~\ref{app:Hqubit}].
For a detailed analysis of the problems at stake, we also considered numerical full diagonalizations using \texttt{LAPACK syevd} libraries~\cite{laug}.
In this, the numerical-matrix representation of the Hamiltonian operator $\hat{H}$ was assembled as a sum of products of finite-matrix representations based on those of the $\hat{b}^{\dagger}$ and $\hat{b}$ operators, with their analytic expressions in the harmonic-oscillator primitive basis, and using different orderings (unordered and normal-ordered). 
Note that the values of the matrix elements of the powers of $ \hat{x} $ and $ \hat{p} $ can be compared to closed formulae, such as those derived for example by Chang~\citep{CHANG2005102}.

\begin{figure}
\resizebox{1\columnwidth}{!}{
\includegraphics[scale=1]{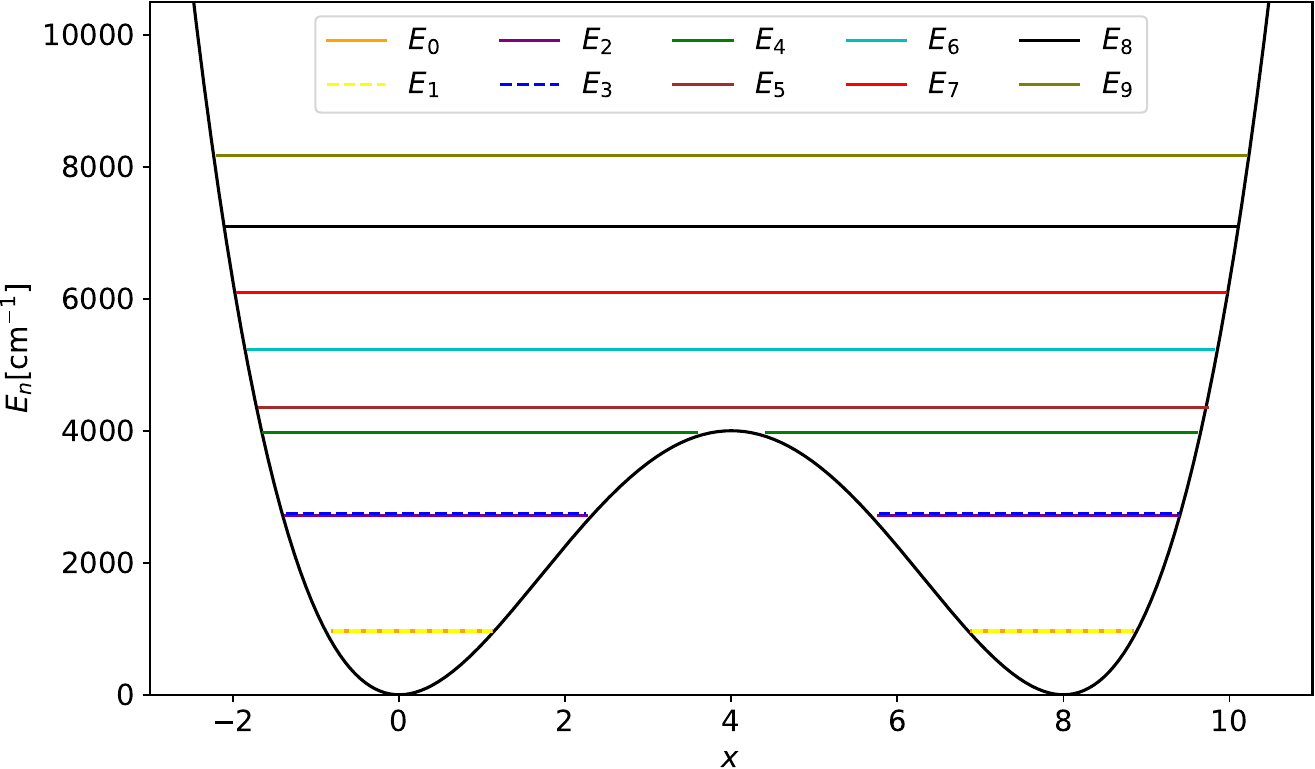}
}
\caption{Double-well potential energy along the $x$ coordinate, together with the first ten eigenvalues, $E_{n}$ (given in cm$^{-1}$).}
\label{fig:potdbW}
\end{figure}

The potential energy of our model, together with the first ten eigenvalues, are shown in Fig. \ref{fig:potdbW}, with an energy scaling factor $\hbar\omega \equiv 2000$ $\text{cm}^{-1}$.
The first two tunneling eigenvalue pairs (0-1; 2-3) are below the transition barrier ($4000$ $\text{cm}^{-1}$).
The following eigenvalue (4) is almost at the barrier.
The next ones are above and start becoming dominated first by a quadratic-potential behaviour, then by a quartic-potential one where the consecutive energy gap keeps increasing gently.

\section{Results and Discussion}\label{sec:resdisc}

In the present section, we provide and discuss illustrations of the numerical effects of the formal aspects that we have mentioned above.

\subsection{Effect of the ordering}

Considering the Hamiltonian given in Eq.~(\ref{eq:hleft}), the ``direct''
expansion of all the monomials of the $\hat{p}$ and $\hat{x}$ operators in terms of $\hat{b}$ and $\hat{b}^\dagger$, according to Eq.~(\ref{eq:canorel}), yields an ``unordered'' representation,
\begin{eqnarray}
\label{eq:hunordered}
&& \hat{H}_{\text{unordered}} = \frac{1}{2}(\hat{b}^{\dagger}\hat{b}+ \hat{b}\hat{b}^{\dagger}) \nonumber \\
&& - \frac{1}{4 \ell \sqrt{2}}(\hat{b}^{3} + \hat{b}^{2}\hat{b}^{\dagger} + \hat{b}\hat{b}^{\dagger}\hat{b} + \hat{b}\hat{b}^{\dagger 2} + \hat{b}^{\dagger}\hat{b}^{2} +\hat{b}^{\dagger}\hat{b} \hat{b}^{\dagger}+  \hat{b}^{\dagger 2}\hat{b} + \hat{b}^{\dagger 3} ) \nonumber \\ 
&& + \frac{1}{32 \ell^{2}}(\hat{b}^{4} + \hat{b}^{3}\hat{b}^{\dagger} + \hat{b}^{2}\hat{b}^{\dagger}\hat{b} + \hat{b}\hat{b}^{\dagger}\hat{b}^{2} + \hat{b}^{\dagger}\hat{b}^{3} + \hat{b}^{2}\hat{b}^{\dagger 2} + \hat{b}^{\dagger}\hat{b}^{2}
\nonumber \\
&& +\hat{b} \hat{b}^{\dagger 2}\hat{b}  + \hat{b} \hat{b}^{\dagger} \hat{b} \hat{b}^{\dagger} 
+\hat{b}^{\dagger}\hat{b}\hat{b}^{\dagger}\hat{b} + \hat{b}^{\dagger}  \hat{b}^{2}\hat{b}^{\dagger} + \hat{b}^{\dagger 2} \hat{b}^{2}+ \hat{b}^{\dagger 3} \hat{b} \nonumber \\
&& + \hat{b}^{\dagger 2}\hat{b}\hat{b}^{\dagger}+ \hat{b}^{\dagger }\hat{b}\hat{b}^{\dagger 2}+ \hat{b}\hat{b}^{\dagger 3}+\hat{b}^{\dagger 4} ) \quad.
 \end{eqnarray}
Quite obviously, it involves all possible unordered products of $\hat{b}^{\dagger}$ with $\hat{b}$ and $\hat{b}$ with  $\hat{b}^{\dagger}$ up to fourth order.
In contrast, getting the expression of the ``ordered'' representation requires some preliminary reshuffling work based on the bosonic commutation relation [Eq.~(\ref{eq:bosocom})] so as to involve only terms with powers of $\hat{b}^{\dagger}$ on the left and powers of $\hat{b}$ on the right according to what is known as Wick's normal order.
In practice, this prescription can make use, as much as possible, of various instances of $\hat{n}=\hat{b}^{\dagger}\hat{b}=\hat{1}+\hat{b}\hat{b}^{\dagger}$, which finally yields
\begin{eqnarray}
\label{eq:hordered}
&&\hat{H}_{\text{ordered}} = \frac{1}{2} \left( 2 \hat{n} + \hat{1} \right) \nonumber \\
&& - \frac{1}{4 \ell \sqrt{2}}(\hat{b}^{3} + 3\hat{n}\hat{b} + 3\hat{b} + 3\hat{b}^{\dagger} + 3\hat{b}^{\dagger}\hat{n} + \hat{b}^{\dagger 3} ) \nonumber \\
 && +\frac{1}{32 \ell^{2}}(\hat{b}^{4} + 4\hat{n}\hat{b}^{2} + 6 \hat{b}^{2} + 6 \hat{b}^{\dagger }\hat{n}\hat{b} + 12 \hat{n}+ 3\hat{1} \nonumber \\
 && +6 \hat{b}^{\dagger 2} + 4\hat{b}^{\dagger 2}\hat{n} +\hat{b}^{\dagger 4} )  \quad.
\end{eqnarray}
The original incentive for usually preferring Wick's normal order within the literature is that only the term involving the identity provides a nonzero contribution to the mean energy of the vacuum state, owing to $\bra{0}\hat{b}^{\dagger} \equiv 0$ and $\hat{b}\ket{0} \equiv 0$. Here, from $\hat{H}_{\text{ordered}}$, we clearly have
\begin{equation}
\expval{\hat{H}}{0}=\frac{1}{2}+\frac{3}{32 \ell^{2}} \quad.
\end{equation}

It must be understood that both expressions of $\hat{H}$ are equivalent and exact, in principle, as long as we assume unlimited access to the infinite harmonic-oscillator basis.
However, in practice (\emph{i.e.}, using a truncated basis of finite size $M$ associated to the projector $\hat{P}_{M} = \sum\limits_{i=0}^{M-1} | i \rangle \langle i |$; see Appendix~\ref{app:Projectors}), the projected variational approximation of $\hat{H}_{\text{unordered}}$, $\hat{P}_{M}\hat{H}_{\text{unordered}}\hat{P}_{M}$, will yield a nonvariational error when one goes as far as replacing within its expression each occurrence of $\hat{b}$ and $\hat{b}^{\dagger}$ by their own projected restrictions, $\hat{b}_{M}=\hat{P}_{M}\hat{b}\hat{P}_{M}$ and $\hat{b}^{\dagger}_{M}=\hat{P}_{M}\hat{b}^{\dagger}\hat{P}_{M}$.
This arises from an incomplete resolution of the identity within products of operators, the effect of which being that the commutation relation satisfied by the projected $\hat{b}_{M}$ and $\hat{b}^{\dagger}_{M}$ is different from the projection of the standard  bosonic one, and rather reads
\begin{equation}
[\hat{b}_{M},\hat{b}^{\dagger}_{M}] = \hat{P}_{M} - | M-1 \rangle M \langle M-1 | \neq \hat{P}_{M} \quad.
\end{equation}

In contrast, doing such a ``brutal'' replacement within the projected variational approximation of $\hat{H}_{\text{ordered}}$, $\hat{P}_{M}\hat{H}_{\text{ordered}}\hat{P}_{M}$, remarkably preserves the variational principle because the potential error due to the incomplete resolution of the identity occurs to cancel out per construction, as proved in Appendix~\ref{app:Projectors}.
Of course, the same holds for their finite matrix representations.
Let us emphasize here that even the harmonic part of the unordered projected operator, $\frac{1}{2}\hat{P}_{M}(\hat{b}^{\dagger}\hat{b}+ \hat{b}\hat{b}^{\dagger})\hat{P}_{M}$, which is approximated as $\frac{1}{2}(\hat{b}^{\dagger}_{M}\hat{b}_{M}+ \hat{b}_{M}\hat{b}^{\dagger}_{M}) $ when considering an incomplete resolution of the identity, corresponds in fact to an artifical system with a spurious eigenvalue, which Buchdahl specifically named the ``truncated harmonic oscillator model'', subject to an abnormal commutation relation.\citep{10.1119/1.1974004} 
In contrast, when doing the same to its ordered counterpart, we get $\hat{b}^{\dagger}_{M}\hat{b}_{M} + \frac{1}{2} \hat{P}_{M} = \hat{n}_{M} + \frac{1}{2} \hat{P}_{M} = \hat{P}_{M}\hat{h}_0\hat{P}_{M}$, and we duly recover the correct variational approximation of $\hat{h}_0=\hat{n}+\frac{1}{2}\hat{1}$ [see Appendix~\ref{app:Projectors}].

It must be understood that the present analysis occurs to be essential if one aims at encoding $\hat{P}_{M}\hat{H}\hat{P}_{M}$ -- within some practical QC algorithmic context -- as a sum of multiple products of $\hat{b}^{\dagger}_{M}$, $\hat{b}_{M}$, and $\hat{n}_{M}$ from the only knowledge of the action of $\hat{b}^{\dagger}_{M}$ given once and for all, for example, in terms of its unary encoding [see Eq.~(\ref{eq:op_unary})] or of its binary encoding [see Eq.~(\ref{eq:Binary_bdag})], which are both recalled in Subsec.~\ref{sec:mappings}. 
Hence, if and only if we stick to Wick's normal order for the ``polyladder'' expansion of $\hat{H}$, we have a convenient way for alleviating the requirement of having to ``qubitize'' beforehand the actions of all the projected representations of the relevant powers of $\hat{p}$ and $\hat{x}$ with respect to the computational basis.
Note that getting $\hat{b}_{M}$ from $\hat{b}^{\dagger}_{M}$ is straightforward, and mapping $\hat{n}_{M}$ is trivial. Of course, similar considerations hold in classical computing if one wants to assemble the restricted Hamiltonian matrix from a limited algebraic subset of finite basic matrices.

In the present case, let us consider what occurs when mapping $\hat{b}^{\dagger}_{M}$ and $\hat{b}_{M}$ to finite strings of Pauli operators and building the Hamiltonian from them.
Using the aforementioned binary encoding~\citep{huang2022qubitization} [see Eq.~(\ref{eq:Binary_bdag})] with three qubits yields 35 terms in the ordered case, for example.
The unordered case also brings 35 terms, but totally different ones [see Appendix~\ref{app:Hqubit}], as expected from the previous discussion.
This confirms that we actually are not representing the same system.

\begin{figure}
\resizebox{1\columnwidth}{!}{
\includegraphics[scale=1]{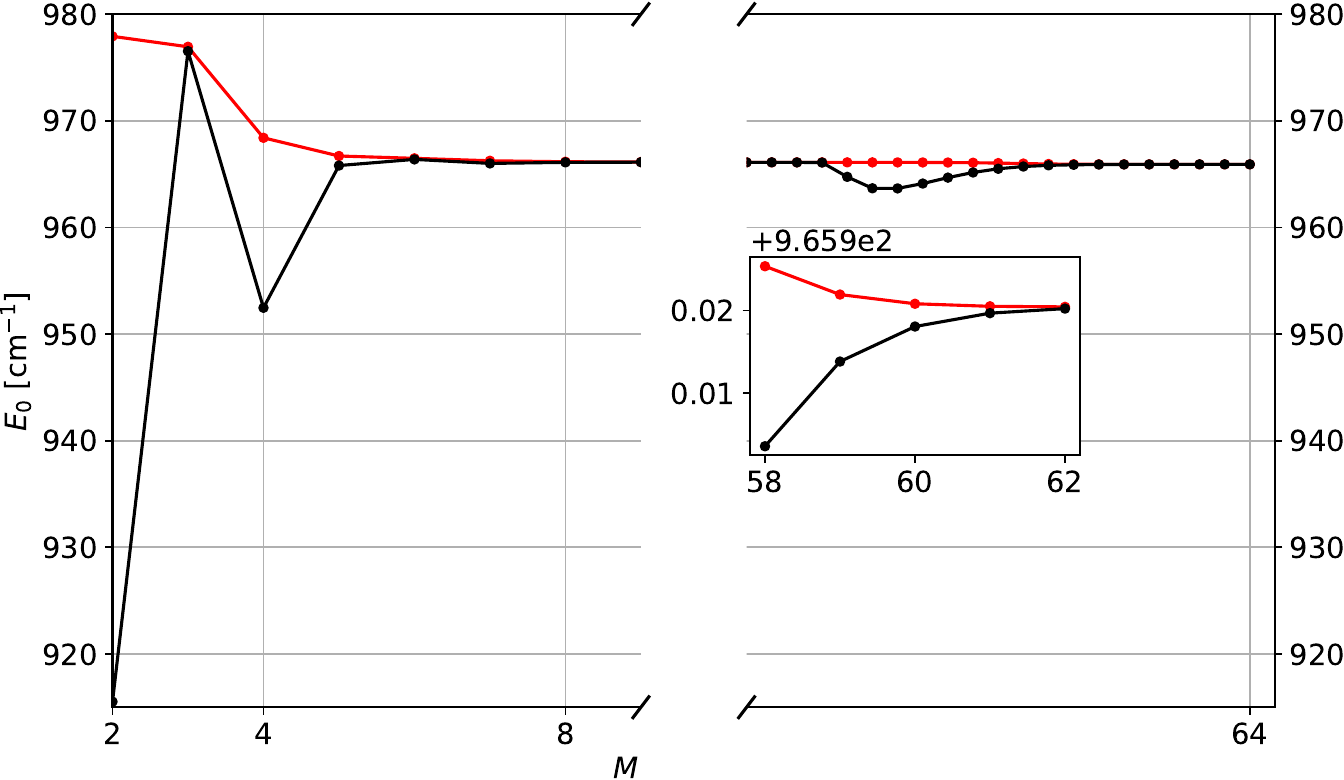}
}
\caption{Convergence behavior of the lowest eigenvalue, $E_{0}$ (given in cm$^{-1}$) when increasing the total number of primitive basis functions, $M$, for both the ordered Hamiltonian ($E_{0}^\text{ordered}$; in red) and the unordered one ($E_{0}^\text{unordered}$; in black).}
\label{fig:E_ordering}
\end{figure}

In what follows, we continue investigating the situation of an incomplete resolution of the identity (\emph{i.e.}, $\hat{b}$ and $\hat{b}^{\dagger}$ are simply replaced beforehand by their restrictions, $\hat{b}_{M}$ and $\hat{b}^{\dagger}_{M}$, within either $\hat{P}_{M}\hat{H}_{\text{unordered}}\hat{P}_{M}$ or $\hat{P}_{M}\hat{H}_{\text{ordered}}\hat{P}_{M}$).
Then, the discrepancy between the ordered or unordered descriptions can be exemplified numerically upon inspecting the behavior of the lowest eigenvalue $E_0$, obtained from exact diagonalization, and its  convergence properties with respect to the size $M$ of the harmonic-oscillator primitive basis.
This is shown in Fig.~\ref{fig:E_ordering}.
The behavior of $E_{0}$ with respect to an increasing value of $M$ is consistent with the variational principle in the ordered case (in red): it decreases smoothly and monotonically down to numerical convergence, which seems to occur visually for $M\approx8$, but, in fact, truly occurs quite later on and only for $M\approx60$ because we are dealing here with fine tunneling conditions.
In contrast, it is evident that, in the unordered case (in black), the behavior of $E_{0}$ with respect to $M$ is not variational at all.
Indeed, several points are lower in energy than the converged solution (first, for $M<8$ -- and quite dramatically for $M=2$ and $M=4$ -- and then again for $M\approx50$).
One could even be mislead at believing that the problem were solved for $M\approx46$ and be content, while a continued increase of the number of basis functions actually keeps making things worse upon creating a new unconverged dip after that.
This is a clear manifestation of the issue related to a wrong ordering whereby the numerical system is polluted by spurious values of the matrix elements related to the abnormal commutation relations induced by an incomplete resolution of the identity.

Hence, even if recasting a nontrivial Hamiltonian according to Wick's normal order may require some extra preliminary work, we stress again that this is a task that should never be circumvented when one want to rely on a finite representation of the ladder operators beforehand.
Our numerical illustration shows that the induced error is \emph{a priori} not benign and should be taken seriously in general situations where its effect is difficult to predict, especially when convergence is entailed not to be rapid, such as in the present model that shows fine tunneling. 

\subsection{Effect of the primitive basis}

Having sorted out the ordering issue, let us stick to a normal-ordered situation and now address the effect of the nature of the primitive basis.
Determining a ``good  basis set'' is essential in any quantum computational approach, and is crucial for QC, since the minimal number of basis functions used for converged simulations is ultimately related to the overall compactness of the quantum circuit, both in terms of width and depth: number of qubits and of operation gates (see for instance Hong {\it et al.}~\cite{hong2022accurate} for the electronic-structure problem).
Hence, we checked here how some minimal change (translation of the coordinate origin, which manifests itself as a unitary change of the underlying primitive basis, generated only by shifting where the center of the harmonic-oscillator basis is located and corresponding to a linear transformation of the ladder operators) could potentially decrease the number of primitive basis functions required for convergence.

So far, we have been considering the primitive basis and its ladder operators as being attached to a harmonic Hamiltonian model with a given origin ($x=0$) centered at the bottom of the left well of the anharmonic Hamiltonian model.
It must be understood that this is a decision to be made in advance:
in general, the primitive basis is chosen to be centered at some relevant stationary point of the potential energy surface (typically, the global minimum or a symmetrical saddle point connecting equivalent minima).
In the case of the one-mode model with a symmetrical double well studied here, there were two obvious possible origin choices: the bottom of one of the two wells (harmonic approximation around a global minimum) or the central point (the top of the transition barrier), intended to better address the left/right symmetry of the problem in a balanced manner.
Thus, we compared these two different choices of origins for the same double-well model,
the latter corresponding to a constant coordinate shift $x \mapsto x+\ell$ with respect to the original model. 
Let us recall here that the choice of origin determines the very definition of $\hat{x}$, hence those of $\hat{b}$ and $\hat{b}^{\dagger}$, which are intrinsically associated to a Gauss-Hermite basis centered at the current choice made for $x=0$ (either the bottom of the left well or the top of the barrier).
In practice, this thus leads to two different formal representations of the Hamiltonian of the same system, the one with the shifted potential energy now being
\begin{eqnarray}
\hat{H}_{\text{center}}&=& \frac{1}{2} \hat{p}^{2} + \frac{\ell^{2}}{8} \hat{1} - \frac{1}{4}\hat{x}^{2}+ \frac{1}{8\ell^2 }\hat{x}^4 \nonumber \\
&=&  \hat{h}_{0} + \frac{\ell^{2}}{8} \hat{1} -\frac{3}{4 }\hat{x}^{2}+ \frac{1}{8\ell^2 }\hat{x}^4 \quad.
\end{eqnarray}

\begin{figure}
\resizebox{1\columnwidth}{!}{
\includegraphics[scale=1]{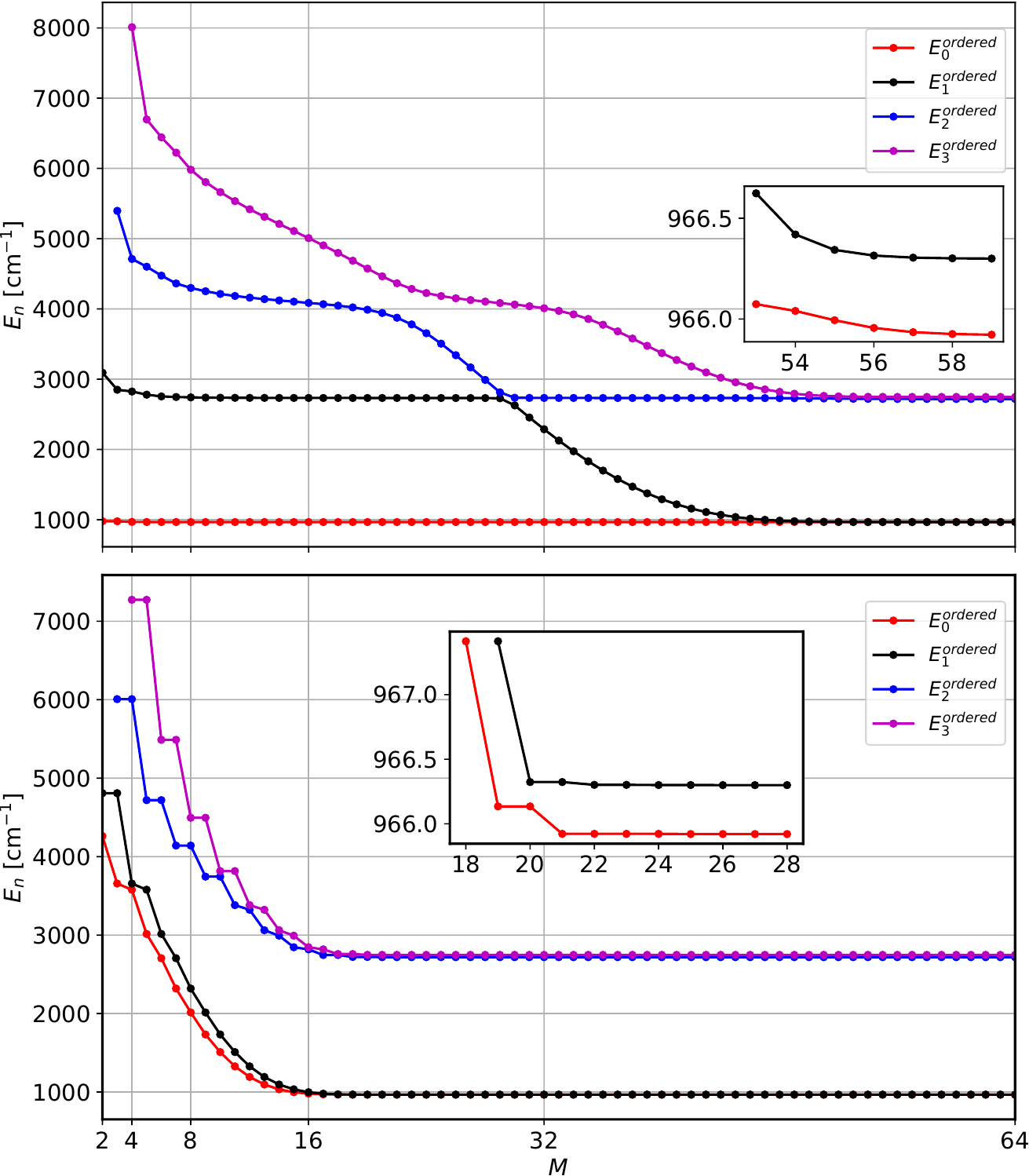}
}
\caption{Behavior of the first four eigenvalues (given in cm$^{-1}$) when increasing the number of basis functions, $M$. The origin of the basis is located at the bottom of the left well, $ \hat{H}_{\text{left}}$ (top panel), or at the top of the barrier, $\hat{H}_{\text{center}}$ (bottom panel).} 
\label{fig:Esupp_shift}
\end{figure}

\newcommand\T{\rule{0pt}{2.6ex}} 
\newcommand\B{\rule[-1.2ex]{0pt}{0pt}} 

 \begin{table*}
        \centering
        \caption{First five eigenvalues in cm$^{-1}$ of the Hamiltonian with the basis centered at the bottom of the left well (top table) and with the basis centered at the top of the central barrier (bottom panel). The index of the eigenvalues goes from 0 to 4 and the digits that differ from the reference values ($M=128$) are highlighted in bold.}
        \label{Eileft_center}
        \begin{tabular}{p{1cm}c c  c c c c c }
           \hline \hline $E_{n,{\rm left}}$ & \multicolumn{6}{ c  }{$M$ \T \B }  \\
            \cline{2-7}
            \T $n$ & 56 & 58 & 60  & 62 \B  & 64 & 128\\
            \cline{1-7}
            0 \T &965.9\textbf{557}  & 965.92\textbf{54} &965.920\textbf{8} & 
           965.920\textbf{5}  \B & 965.9204 &  965.9204\\
            1 \T& 966.\textbf{3429} & 966.\textbf{3046} &  966.299\textbf{9 } & 966.299\textbf{6} \B & 966.2995 & 966.2995\\
            2 \T & 2718.\textbf{7273} & 2718.\textbf{4283} &  2718.37\textbf{71} &  2718.371\textbf{5}  \B & 2718.371\textbf{3} & 2718.3712\\
            3 \T & 2747.\textbf{6837} & 2747.\textbf{3439} & 2747.28\textbf{71} & 2747.280\textbf{9} & 2747.2806 \B & 2747.2806\\
            4 \T & 3967.\textbf{8987} & 3967.\textbf{1728} &3967.0\textbf{283}  & 3967.00\textbf{92}\B & 3967.0078 & 3967.0078\\
            \hline \hline \\
            \hline \hline $E_{n,{\rm center}}$ & \multicolumn{6}{ c  }{$M$ \T \B }  \\
            \cline{2-7}
             \T $n$ & 24 & 26 & 28  & 30 \B  & 32 & 128\\
            \cline{1-7}
            0 \T &965.92\textbf{21}  & 965.920\textbf{6} & 965.9204 & 
           965.9204   \B& 965.9204 & 965.9204\\
            1 \T & 966.\textbf{3002} & 966.299\textbf{6} & 966.2995 & 966.2995\B& 966.2995 & 966.2995\\
            2 \T & 2718.3\textbf{813} & 2718.37\textbf{34} & 2718.371\textbf{3} &2718.371\textbf{3}\B& 2718.3712 & 2718.3712\\
            3 \T & 2747.28\textbf{76} & 2747.28\textbf{10} & 2747.2806 & 2747.2806 \B& 2747.2806 & 2747.2806\\
            4 \T & 3967.0\textbf{231} & 3967.0\textbf{146} & 3967.00\textbf{80} & 3967.007\textbf{9} \B& 3967.0078 & 3967.0078 \\
            \hline \hline
        \end{tabular}
\end{table*}

In the top panel of Fig. \ref{fig:Esupp_shift}, which is associated to $\hat{H} = \hat{H}_{\text{left}}$ [see Eq.~(\ref{eq:hleft})], and as already mentioned, we observe that the lowest eigenvalue, $E_0$, seems to be converging fast from visual inspection; however, the subtle effect of the fine tunneling splitting with $E_1$ ($0.38~\text{cm}^{-1}$) appears only for $M\approx60$ basis functions, thus highlighting the strong anharmonicity of this model.
Interestingly enough, one notices that the curves of $E_1$ and $E_2$ cross at $M\approx30$.
On the left side, where the numerical effect of the tunneling has not set off yet, $E_1$ is still dominated by the approximate harmonic description of the left well and its first excited state: it remains around $\hbar \omega (1 + \frac{1}{2}) \approx 3000~\textrm{cm}^{-1}$.
On the right side, $E_1$ finally starts going down variationally to the energy of the upper state of the first tuneling pair with $E_0$ for the lower state ($\approx 1000~\textrm{cm}^{-1}$) while $E_2$ replaces it and duly converges to the energy of the lower state of the second tuneling pair, together with $E_3$ for the upper state ($\approx 3000~\textrm{cm}^{-1}$).
In contrast, in the bottom panel of Fig. \ref{fig:Esupp_shift}, we see that we converge both pairs of quasidegenerate eigenvalues (0-1; 2-3) faster: the tunneling splitting is reached earlier, for $M\approx22$ basis functions. 
The detailed convergence of the first five eigenvalues is given in Table \ref{Eileft_center}.
The reference value for numerically exact convergence is taken as $M=128$.
While the ``left'' model requires up to $M=64$ for converging four decimal places in $\textrm{cm}^{-1}$, we observe that the ``center'' model only requires $M=32$ within the same convergence threshold.
Of course, we could be less demanding (only two or three decimal places), but the numerical behavior is in fact almost the same within this range.

\begin{figure*}
\resizebox{2\columnwidth}{!}{
\includegraphics[scale=1]{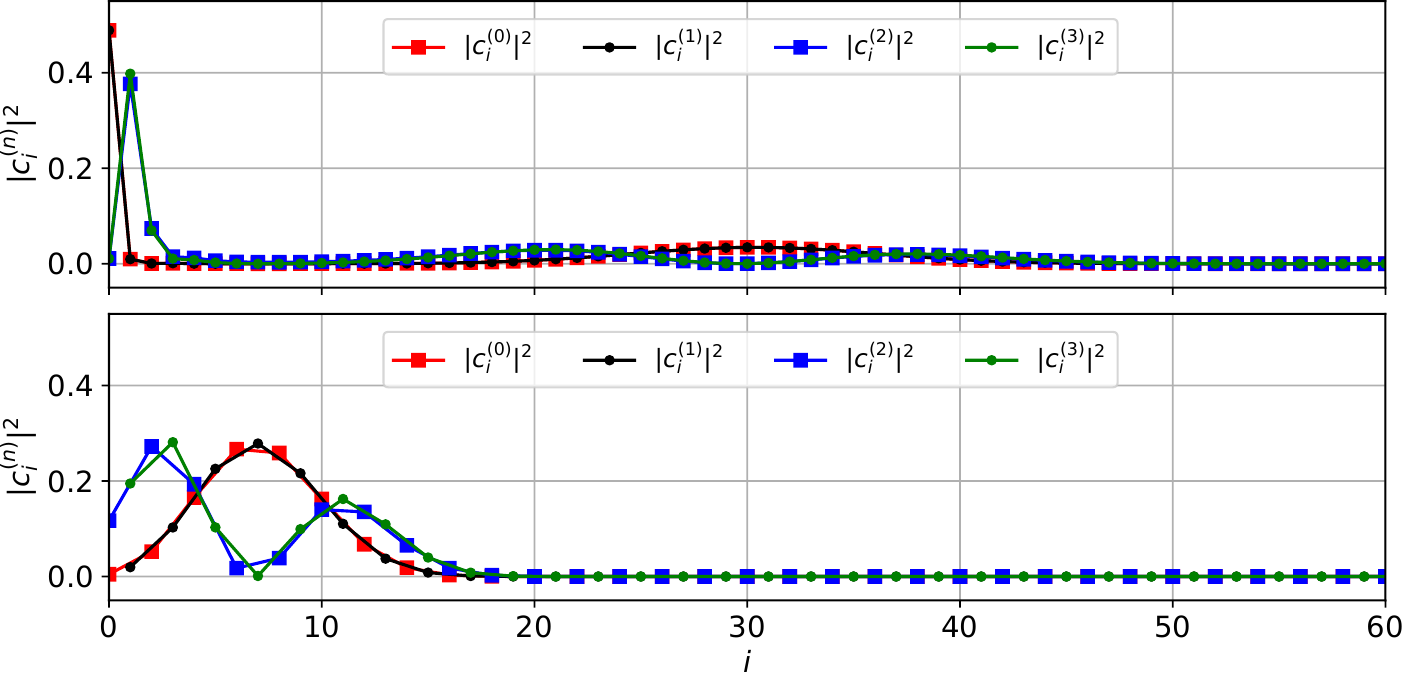}
}
\caption{Behavior of the weights at convergence of the basis functions, $i$, with respect to the first four eigenstates. The origin of the basis is either at the bottom of the left well, $ \hat{H}_{\text{left}}$ (top panel), or at the top of the barrier, $ \hat{H}_{\text{center}}$ (bottom panel).}
\label{fig:weights}
\end{figure*}

Then, analysing the convergence behavior of the weights of the basis functions (index $i$) with respect to the lowest-energy eigenvectors (index $n$), $ w_{i}^{(n)} = | c_{i}^{(n)}|^{2} = |\langle i | \psi^{(n)} \rangle|^{2}$, for the two models is illuminating
and is shown in Fig.~\ref{fig:weights}.
In the case where the basis is centered at the left well, most of the dominant contributions are recovered early on with a handful of basis functions.
In particular, the ground ($n=0$) and first-excited ($n=1$) states of the first tunneling pair (about the normalized plus and minus combinations of the Gaussian functions of the left and right wells) are almost half-described by the single Gaussian function ($i=0$) of the left well. 
Yet, because of fine tunneling, we have to wait for a while until we obtain the effect of the many detailed contributions of highly-excited basis functions, so as to reconstruct symmetrically the Gaussian contribution in the right well from the excited states of the left well (typically peaking around $i \approx 30$ but with quite a large spread).
As already mentioned, such a case is not trivial as regards convergence properties, which is why it was chosen as our focus here.
However, such a behavior is not observed in the case where the basis is centered at the top of the barrier, for which we need to account for some early and large contributions, not beyond around $i \approx 20$).
Hence, the latter case shows much better convergence properties.
In addition, let us remark that this complies with symmetry considerations: even/odd eigenstates only expand over even/odd primitive states.

As a final word, it is interesting to analyse which type of Hamiltonian representation is bound to be more efficient numerically for a simulation based on a quantum algorithm.
To this end, one can compare the 1-norm of the qubit Hamiltonians, computed as the sum of the absolute values of the coefficients associated to the Pauli-string expansion of $\hat{H} = \sum\limits_{i} \lambda_{i} \hat{P}_{i}$ as
\begin{equation}
\lambda = \sum \limits_{i} |\lambda_{i}| \quad.
\end{equation}
Indeed, the complexity of quantum algorithms typically scales with the 1-norm,
like the number of measurements in variational quantum algorithms~\cite{wecker2014gate}
and in the simulation of Hamiltonians using qubitization techniques~\citep{PhysRevResearch.3.033055,lee2021even}.
Changing the basis can have some significant impact on this quantity, as shown in Ref.~\cite{PhysRevResearch.3.033127} for the electronic-structure problem using localized orbitals.
Similar analyses are thus as much relevant for the vibrational-structure problem.

\begin{figure}
\resizebox{1\columnwidth}{!}{
\includegraphics[scale=1]{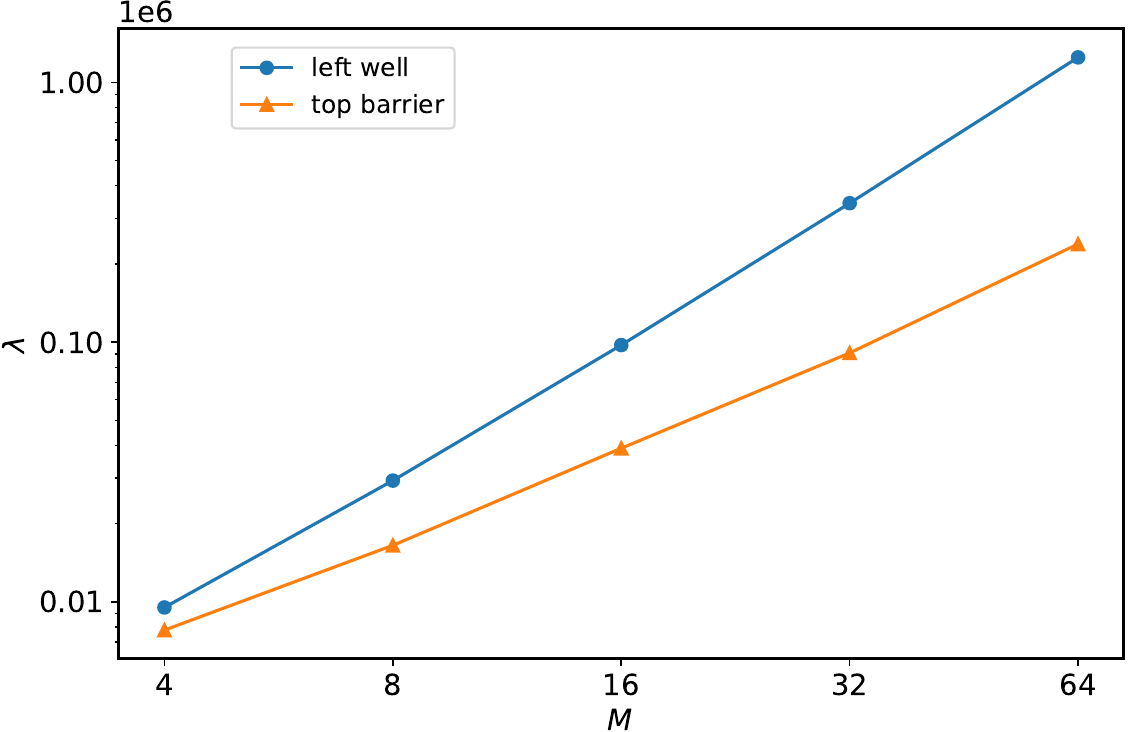}
}
\caption{Behavior (log-log scale) of the 1-norm $\lambda$ of the Hamiltonians with the origin of the basis centered at the left-well bottom in blue or at the barrier top in orange, with respect to the number of basis functions $M$.}
\label{fig:1_norm}
\end{figure}

As readily seen in Fig.~\ref{fig:1_norm}, the slope of the 1-norms associated to the left-basis and center-basis are $\alpha = 1.76 $ and $ \alpha =1.23 $, respectively (in log-log scale).
In addition, as already mentioned, and according to Fig.~\ref{fig:Esupp_shift}, only 5 qubits (32 basis functions within the binary encoding) are required to converge the tunneling splitting when the basis is centered at the top of the barrier, while 6 qubits (64 basis functions) would be required when the basis is centered at the bottom of the left well.
The 1-norm for these two cases differs by one order of magnitude, in favor of the basis centered at the top of the barrier, which should therefore be preferred for a quantum simulation of the one-dimensional vibrational model investigated in this work.

Herein, we have considered two choices of primitive bases, motivated by the topographical features of the potential energy surface, and compared their consequences within a QC context.
In contrast with electronic-structure problems whereby universal and transferable atomic-orbital-like primitive bases have been designed, benchmarked, and pre-optimized for decades, according to their abilities to best reproduce experimental observables, it must be undertood that such an aspect is too much system-dependant for vibrational-structure problems.
Defining first a relevant system of internal coordinates and then choosing an optimal primitive basis can thus be viewed as the heuristic art of quantum dynamics. 

Yet, such a question still remains open, especially within a QC context.
In order to get further, we could envision some rational strategies for optimally contracting the variational basis.
In analogy to the use of local orbitals to reduce the 1-norm in the electronic-structure problem~\cite{PhysRevResearch.3.033127}, using local modes could also reduce the 1-norm of vibrational-structure Hamiltonians compared to the delocalized VSCF modals. 
Alternatively, there may some room to explore in analogy to the use of Daubechies wavelet molecular orbitals in electronic-structure theory rather than Gaussian-type orbitals~\cite{hong2022accurate}, showing that fewer quantum resources could be used to solve the same problem with similar accuracy.
Such studies are left for future work.

\section{Conclusions and outlook}

In this paper, we highlighted the fundamental issues that occur when the primitive basis is truncated in practice.
Beyond the evident variational convergence properties with respect to the size of the finite basis, we also explored the less-known effect of various orderings of products of bosonic operators when defining the algebraic Hamiltonian representation to be mapped ultimately to qubits via unary or binary encodings.
This was illustrated numerically on a one-dimensional Hamiltonian model corresponding to a nontrivial double-well potential showing fine tunneling and no upper bound, so as to drag attention onto the actual numerical effect of such a formal error, beyond the variational one, that may arise when simulating a vibrational problem on a quantum computer using a finite basis for a bosonic system with an infinite Hilbert space. 
We thus confirmed that using the so-called normal order is essential indeed and should never be circumvented, even if this requires extra preliminary work. 

We also addressed the practical question of the optimal choice of a good basis set as regards the efficiency of variational convergence.
In the case of a strongly anharmonic system, the first decision to be made is the location of the center of the primitive basis functions with respect to the origin of the potential energy function.
This was illustrated here by a detailed analysis of convergence (eigenvalues and eigenvectors) comforted by an investigation of the 1-norm, which relates to the efficiency of simulating vibrational-structure problems with QC algorithms. Further benchmarks are then required to draw general conclusions on the choice of an optimal basis set to be used in QC simulations of vibrational-structure theory.

We hope that the present work may benefit to researchers already doing vibrational structure with classical computing but aiming at exploring the capabilities offered by QC, as well as to researchers already doing QC for electronic structure but interested in generalizing their algorithms to vibrational structure.

\section*{Data availability}

No new data regarding actual systems were generated or analyzed in support of this study.
The defining parameters of our models and the softwares and libraries that were used in our computations are all freely available and duly listed in the published article.

\section*{Acknowledgments}

We acknowledge the Institute for Quantum Technologies in Occitanie for its support through the PhD funding program AMI CMA QuantEDU-France (project `VibElQuant').
We warmly thank Saad Yalouz for regular discussions and relevant scientific input as regards the relevance of the present study.

\appendix

\section{Harmonic ladder operators and their properties}\label{app:ladder}

\subsection{The harmonic oscillator prototype for a bosonic mode}

Let us consider a generic, spinless, one-dimensional/one-particle quantum system.
In what follows, we shall use nondimensionalized units (where, in particular, $\hbar=1$).
In the absence of any external field, such a system is determined by a time-independent and spin-free Hermitian Hamiltonian, $\hat{H}$, expressed as an analytic classical functional of its Hermitian position and momentum operators, $\hat{x}$ and $\hat{p}$.
The latter obey canonical commutation rules (responsible of the fundamental Heisenberg uncertainty principle),
\begin{equation}
\label{eq:cancom}
[\hat{x}, \hat{p}] = i\hat{1} \quad, 
\end{equation}
where $\hat{1}$ is the identity operator of the system Hilbert space.
Note that we implicitly assume vanishing boundary conditions, such that being self-adjoint here boils down to being Hermitian.
Then, we can define the following non-Hermitian operators,
 \begin{equation}
    \hat{b} = \frac{1}{\sqrt{2}}(\hat{x}  +i \hat{p}) \quad, \hspace{1cm} \hat{b}^{\dagger} = \frac{1}{\sqrt{2}}(\hat{x}  -i \hat{p}) \quad,
\end{equation}
such that
\begin{equation}
\label{eq:xpbb}
    \hat{x} = \frac{1}{\sqrt{2}}(\hat{b}+\hat{b}^{\dagger}) \quad, \hspace{1cm} \hat{p} = \frac{1}{i\sqrt{2}}(\hat{b}-\hat{b}^{\dagger}) \quad.
\end{equation}
The latter are Hermitian; they obviously correspond to complex-conjugate phase-space variables and are often invoked within the context of coherent states (eigenstates of $\hat{b}$).
The corresponding bosonic commutation rule follows from Eq.~(\ref{eq:cancom}) and reads
\begin{equation}
\label{eq:com2nd}
[\hat{b}, \hat{b}^{\dagger}] = \hat{1} \quad.
\end{equation}

Now, it occurs that $\hat{b}$ and $\hat{b}^{\dagger}$ act as the ladder operators of the quantum harmonic-oscillator model with respect to its Fock space.
This provides a natural primitive basis set in which the actions of the various ladder operators and their products can be made explicit.
The nondimensionalized harmonic Hamiltonian model reads
\begin{equation}
\label{eq:HOxp}
 \hat{h}_{0} = \frac{1}{2}(\hat{p}^{2}+\hat{x}^{2}) \quad,
\end{equation}
with natural harmonic units (energy measured in terms of $\hbar \omega $, length in terms of $\sqrt{\frac{\hbar}{m\omega}}$, and momentum in terms of $\sqrt{\hbar m \omega}$; this implies to take $\hbar=1$, $m=1$, and $\omega=1$ in practice).
The eigenfunctions of the quantum harmonic-oscillator model in position representation are the Gauss-Hermite functions,
\begin{equation}
\langle x | n \rangle = \phi_{n}(x) = \frac{1}{\sqrt{\pi^{1/2} 2^{n}n!}} e^{-\frac{x^{2}}{2}} \mathcal{H}_{n}(x) \quad,
\end{equation}
where $ \mathcal{H}_{n} $ is the physicists' Hermite polynomial of degree $n$, for any $ n \in \{ 0, 1, 2 ... , \infty \}$.

As regards the harmonic basis set, $\{|n\rangle\}_{n=0}^\infty$, we have the following properties with respect to $\hat{b}$ and $\hat{b}^{\dagger}$, for any nonnegative integer $n$,
\begin{equation}
\hat{b}^{\dagger} |n \rangle = \sqrt{n+1} |n+1 \rangle \quad,
\end{equation}
\begin{equation}
\hat{b} |n \rangle = \sqrt{n} |n-1 \rangle \quad,
\end{equation}
augmented for consistency by the terminating ``quantum vacuum'' condition,
\begin{equation}
\hat{b} |0 \rangle \equiv 0 \quad.
\end{equation}
Further, we have the number operator, $\hat{n} = \hat{b}^{\dagger}\hat{b}$, which brings
\begin{equation}
\hat{n} |n \rangle = n |n \rangle \quad.
\end{equation}
This allows us to establish a one-to-one correspondence between the Hilbert space of the vibrational system and the Fock space of the bosonic mode upon considering the index of the harmonic eigenvector, $n$, as the number of bosons in the harmonic mode.

It must be stressed here that the bosonic commutation relations still hold whether the actual Hamiltonian $\hat{H}$ is the harmonic one, $\hat{h}_{0}$, or something more complicated.
Hence, for an anharmonic Hamiltonian, the harmonic ladder operators are no longer the bosonic ladder operators of the eigenbasis but only the bosonic ladder operators attached to the primitive harmonic-oscillator (Gauss-Hermite) basis $\{|n\rangle\}_{n=0}^\infty$.
Yet, knowing the actions of $\hat{b}$ and $\hat{b}^{\dagger}$ with respect to the primitive basis allows us to determine conveniently the actions of $\hat{x}$ and $\hat{p}$, and of any of their powers involved in a Hamiltonian that can be written as an analytic functional of those. 
In particular, in the present work, we have explored a situation where the kinetic-energy operator remains $\frac{1}{2}\hat{p}^{2}$ (rectilinear coordinate) but where the potential-energy operator is a confining lower-bounded quartic polynomial of $\hat{x}$.

Now, expressing $ \hat{p}^{2} $ and $ \hat{x}^{2}$ directly in terms of the harmonic ladder operators [direct
expansion from Eq. (\ref{eq:xpbb})] yields 
\begin{equation}
 \hat{h}_{0} = \frac{1}{2} \left( \hat{b}^{\dagger}\hat{b} + \hat{b} \hat{b}^{\dagger} \right) \quad.
\label{eq:directh0}
\end{equation}
While such an expression seems perfectly valid at first sight, it brings potential sources of error when implementing it in practice for a truncated basis set (this will be discussed in the next subsection). 
From the commutation rule set in Eq. (\ref{eq:com2nd}), we can devise two alternative, and \textit{a priori} equivalent, expressions of $\hat{h}_{0}$,
\begin{equation}
 \hat{h}_{0} = \frac{1}{2} \left( 2 \hat{b}^{\dagger}\hat{b} + \hat{1} \right) = \frac{1}{2} \left( 2 \hat{b}\hat{b}^{\dagger} - \hat{1} \right) \quad.
\label{eq:orderh0}
\end{equation}
Within the second-quantized context of Wick's theorem \cite{Wick}, the first expression is known as ``normal ordered'', while the second as ``antinormal ordered''.
Again, they are both equally valid in principle, but the ``normal ordered'' one is often preferred in the literature because it brings a single contribution for the vacuum state, due to the identity term, as exemplified in the main text (here, $\expval{\hat{h}_{0}}{0}=\frac{1}{2})$.
In any case, using the definition of the bosonic counting (quantum number) operator, this brings the well-known second-quantized form of the nondimensionalized harmonic Hamiltonian model,
\begin{equation}
 \hat{h}_{0} = \frac{1}{2} \left( 2 \hat{b}^{\dagger}\hat{b} + \hat{1} \right) = \hat{n} + \frac{1}{2} \hat{1} \quad.
\end{equation}
Obviously, $\hat{h}_{0}$ and $\hat{n}$ commute and share the same set of eigenvectors, which satisfy
\begin{equation}
\hat{h}_{0} |n \rangle = \left(n+\frac{1}{2}\right) |n \rangle \quad.
\end{equation}

\subsection{Ordering issues for finite basis sets}\label{app:Projectors}

The harmonic-oscillator basis, defined above as $\{|n\rangle\}_{n=0}^\infty$, and attached to $\hat{b}$ and $\hat{b}^{\dagger}$ as its ladder operators, is discrete but infinite.
However, under practical instances, it will have to be truncated so as to provide finite matrix representations of relevant operators. As recalled below, there is a fundamental issue that concerns the truncated representation of a product of two operators with a finite matrix and whether or not it can be identified to the product of two such finite matrices.

We shall refer to $M>0$ as the finite number of basis functions under computational circumstances.
Restricted operators can be expressed upon using the idempotent projector: $\hat{P}_{M} = \sum\limits_{i=0}^{M-1} | i \rangle \langle i |$ where $\hat{P}_{M}^2=\hat{P}_{M}$.
The truncated matrix representations of the operators will have an $(M \times M)$-dimension. Formally, the corresponding projected ladder operators read
\begin{equation}
\hat{b}_{M} = \hat{P}_{M}\hat{b}\hat{P}_{M} \quad, \quad \hat{b}_{M}^{\dagger} = \hat{P}_{M}\hat{b}^{\dagger}\hat{P}_{M} \quad.
\end{equation}
The problem at stake here is that 
\begin{equation}
[\hat{b}_{M},\hat{b}^{\dagger}_{M}] \neq \hat{P}_{M} \quad.
\end{equation}
Note that $\hat{P}_{M}$ can be viewed as the projected representation of the identity operator within the finite subspace: $\hat{P}_{M}=\hat{P}_{M}\hat{1}\hat{P}_{M}\equiv\hat{1}_{M}$.
Hence, we can no longer use the standard bosonic commutation relation to switch safely from some choice of ordering to another within the truncated subspace.

The proof is simple and follows. 
First, upon expanding the projectors involved within $\hat{b}_{M}^{\dagger}\hat{b}_{M}$, we get

\begin{eqnarray}
&&\hat{P}_{M}\hat{b}^{\dagger}\hat{P}_{M}\hat{b} \hat{P}_{M} =  \sum\limits_{i=0}^{M-1} \sum\limits_{j=0}^{M-1} \sum\limits_{k=0}^{M-1} |i \rangle \langle i | \hat{b}^{\dagger} |j \rangle \langle j | \hat{b} |k \rangle \langle k |  \nonumber \\
&=& \sum\limits_{i=0}^{M-1} \sum\limits_{j=0}^{M-1} \sum\limits_{k=0}^{M-1} |i \rangle \langle i  |j +1 \rangle (j+1) \langle j +1 |k \rangle \langle k | \nonumber \\
&= & \hat{P}_{M}| 0 \rangle 0 \langle 0 |\hat{P}_{M} + \sum\limits_{i=0}^{M-1} \sum\limits_{l=1}^{M-1} \sum\limits_{k=0}^{M-1} |i \rangle \langle i  |l \rangle l \langle l |k \rangle \langle k | \nonumber \\
&& +  \hat{P}_{M}  |M\rangle M\langle M | \hat{P}_{M} \nonumber \\
&= & \hat{P}_{M}\hat{n}\hat{P}_{M} \nonumber \\
&=& \hat{P}_{M}\hat{b}^{\dagger}\hat{b}\hat{P}_{M} \quad,
\end{eqnarray}
where we used $ l = j+1 $, the fact that incorporating $| 0 \rangle 0 \langle 0 |$ into the sum is merely adding up zero, and the property that $\hat{P}_{M}  |M\rangle \equiv 0$ (orthogonality of the subspace with respect to its complement).
Second, for $\hat{b}_{M}\hat{b}_{M}^{\dagger}$, we have
\begin{eqnarray}
&&\hat{P}_{M}\hat{b}\hat{P}_{M}\hat{b}^{\dagger} \hat{P}_{M} = \sum\limits_{i=0}^{M-1} \sum\limits_{j=0}^{M-1} \sum\limits_{k=0}^{M-1} |i \rangle \langle i | \hat{b} |j \rangle \langle j | \hat{b}^{\dagger} |k \rangle \langle k | \nonumber \\
& =& 0\hat{P}_{M} + \sum\limits_{i=0}^{M-1} \sum\limits_{j=1}^{M-1} \sum\limits_{k=0}^{M-1} |i \rangle \langle i  |j -1 \rangle j \langle j-1 |k \rangle \langle k | \nonumber \\ 
&= & \sum\limits_{i=0}^{M-1} \sum\limits_{l=0}^{M-2} \sum\limits_{k=0}^{M-1} |i \rangle \langle i  |l \rangle (l+1) \langle l |k \rangle \langle k | \nonumber \\  
&= & \sum\limits_{i=0}^{M-1} \sum\limits_{l=0}^{M-1} \sum\limits_{k=0}^{M-1} |i \rangle \langle i  |l \rangle (l+1) \langle l |k \rangle \langle k | \nonumber \\
&&- \sum\limits_{i=0}^{M-1}\sum\limits_{k=0}^{M-1} |i \rangle \langle i | M-1 \rangle M \langle M-1 |k \rangle \langle k  |\nonumber \\ 
&= & \hat{P}_{M}(\hat{n} + \hat{1})\hat{P}_{M} - \hat{P}_{M}| M-1 \rangle M \langle M-1 |\hat{P}_{M} \nonumber \\
&= & \hat{P}_{M}\hat{b}\hat{b}^{\dagger}\hat{P}_{M} - | M-1 \rangle M \langle M-1 | \quad,
\end{eqnarray}
where we used the vacuum condition and $ l = j-1 $.
The last term is the culprit: it is now nonzero since it belongs to the truncated subspace. 
It follows that
\begin{equation}
[\hat{b}_{M},\hat{b}^{\dagger}_{M}] = \hat{P}_{M} - | M-1 \rangle M \langle M-1 | \neq \hat{P}_{M} \quad.
\end{equation}
For illustrating purposes, let us consider a matrix representation, with  $\textbf{b}^{\dagger}_{M} \textbf{b}_{M}$ and  $\textbf{b}_{M}\textbf{b}^{\dagger}_{M}$ for $M=4$, the elements of which being analytic integrals that are known explicitly. 
Their normal-ordered and antinormal-ordered products are, respectively,
\begin{center}
$ \textbf{b}^{\dagger}_{4} \textbf{b}_{4} = \begin{bmatrix}
0 & 0 & 0 & 0\\
0 & 1 & 0& 0\\
0 & 0 & 2 & 0\\
0 & 0 & 0 & 3
\end{bmatrix}
= \textbf{n}_{4}$
\hspace{0.5\baselineskip} but \hspace{0.5\baselineskip}
$ \textbf{b}_{4}\textbf{b}^{\dagger}_{4} =  \begin{bmatrix}
1 & 0 & 0 & 0\\
0 & 2 & 0& 0\\
0 & 0 & 3 & 0\\
0 & 0 & 0 & 0
\end{bmatrix}
= \textbf{n}_{4}+\textbf{1}_{4} - \begin{bmatrix}
0 & 0 & 0 & 0\\
0 & 0 & 0& 0\\
0 & 0 & 0 & 0\\
0 & 0 & 0 & 4
\end{bmatrix}$
\quad.
\end{center}
This also modifies consistently the canonical commutation relation between the projected position and momentum operators, 
\begin{equation}
[\hat{x}_{M},\hat{p}_{M}] = i(\hat{P}_{M} - | M-1 \rangle M \langle M-1 |) \neq i\hat{P}_{M} \quad,
\end{equation}
where 
\begin{equation}
\hat{x}_{M} = \hat{P}_{M}\hat{x}\hat{P}_{M} \quad, \quad \hat{p}_{M} = \hat{P}_{M}\hat{p}\hat{P}_{M} \quad.
\end{equation}
Such a predicament was pointed out by Buchdahl as early as 1967 within the general framework of the so-called ``truncated harmonic oscillator model''. \citep{10.1119/1.1974004}
There, it was noticed that the truncated harmonic Hamiltonian defined as 
\begin{equation}
 \hat{h}_{0M} = \frac{1}{2}(\hat{p}_M^{2}+\hat{x}_M^{2}) \quad,
\end{equation}
together with its abnormal canonical commutation relation (see above), could be reconsidered formally as sharing the Lie algebraic properties of a finite angular-momentum Hamiltonian.
This was revisited in 2003 within a QC context by Somma \emph{et al.}~\citep{SommaIJQI} -- and further explored algebraically by Batista and Ortiz in 2004~\citep{Batista01012004} -- who proposed an alternative formulation based on ladder operators only,
\begin{equation}
[\hat{b}_{M},\hat{b}^{\dagger}_{M}] = \hat{1}_{M} - M \frac{(\hat{b}_M^\dagger)^{M-1} }{\sqrt{(M-1)!}} \frac{(\hat{b}_M)^{M-1} }{\sqrt{(M-1)!}} \quad,
\end{equation}
together with the assumption that $(\hat{b}_M^\dagger)^{M} = \hat{0}$.
The equivalence may seem intuitive from
Eq.~(\ref{eq:dyadharm}) but requires to carefully check that we duly get rid of the presence of  $\ket{0}\bra{0}$ in the middle of the expansion of $\ket{M-1}\bra{M-1}$. This can be proved pedestrianly upon comparing their respective matrix elements, with left and right indices ranging from $0$ to $M-1$, to show that only the bottom-right term survives.

Such a formal correspondence between a truncated oscillator and a finite angular-momentum system resonates directly with the seminal works of Jordan~\citep{Jordan1935} in 1935 and Schwinger~\citep{Schwinger1952} in 1952 about the formal mapping to be made between the $\mathfrak{su}(2)$ Lie algebra of angular momenta ($\ket{j,m}$ states) and the bosonic Weyl algebra of two uncoupled truncated harmonic modes ($\ket{n_1,n_2}$ states), which was explored somewhat complementarily in 1940 by Holstein and Primakoff~\citep{PhysRev.58.1098}, upon realizing that fixing $j$ to some nonnegative integer or half-integer value according to the irreducible representations of the $\mathrm{SU}(2)$ Lie group, and letting $m$ vary among $2j+1$ values such that $-j \leq -j+1 \leq \dots \leq m \leq \dots \leq j-1 \leq j$, reduces the system to a single truncated harmonic mode, thanks to the conservation -- as a constant of motion -- of the so-called `polyad': $j=n_1+n_2$~\citep{PhysRevA.59.64}.
The literature on this subject is vast~\citep{PhysRevA.103.042209, PhysRevLett.78.578,PhysRevA.59.64, 10.1063/1.438598,Batista01012004},
and an exhaustive survey is beyond the scope of the present work; let us, however, mention some recent work~\citep{Sawaya_2020} in 2020, reconsidering within a QC context the MMST-like mapping~\citep{PhysRevLett.78.578,PhysRevA.59.64,10.1063/1.438598} between a finite $(2j+1)$-state bosonic-like Hamiltonian and a spin-$j$ angular-momentum Hamiltonian, together with their relations via the $\mathfrak{su}(2)$ Lie algebra, the $j$-representations of the $\mathrm{SU}(2)$ Lie group, and the generators of the $\mathrm{SU}(2j+1)$ Lie group~\citep{Batista01012004}.

\begin{figure}
\resizebox{0.8\columnwidth}{!}{
\includegraphics[scale=1]{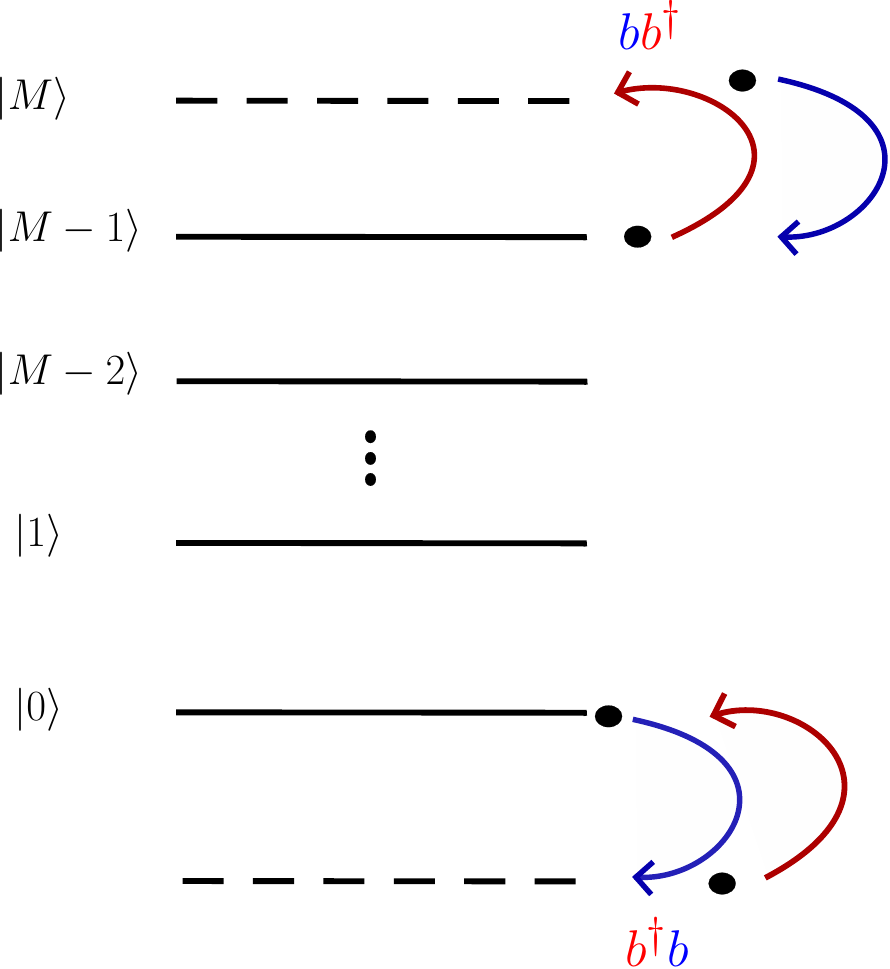}
}
\caption{Schematic description of the action of bosonic ladder operator products with respect to a truncated harmonic-oscillator basis.}
\label{fig:ordering}
\end{figure}

Again, and as regards the present work, it should be understood that -- on practical terms -- the problem is somewhat simpler and can be viewed essentially as an incomplete resolution of the identity (also known as the closure relation) for an infinite Hilbert space when inserting a finite projector within a product of two operators. 
We can summarize the second-quantized description of a truncated bosonic mode with a finite set of ladder operators according to the scheme presented in Fig.~\ref{fig:ordering}.

Whether the basis is truncated or not,
$\hat{b} $ is always lower-bounded by the $ |0 \rangle $ state (the ``physical vacuum'').
The basis truncation also implies that applying the creation operator to the last state of the truncated basis leads to a state that is not included in the truncated subspace.
Thus, this creates an upper bound to the number of bosons that can be created in the mode, and a sort of extra ``numerical vacuum'' from above.
As shown above, a practical solution is found when invoking the ``normal ordered'' product, $\hat{b}^{\dagger}_{M}\hat{b}_{M}=\hat{P}_{M}\hat{b}^{\dagger}\hat{P}_{M}\hat{b}\hat{P}_{M}=\hat{P}_{M}\hat{b}^{\dagger}\hat{b}\hat{P}_{M}=\hat{P}_{M}\hat{n}\hat{P}_{M}$, while using any type of ``unordered'' product causes troubles.
Such a situation is general and applies to monomials of higher degree.

Hence, while it is well-known that the historical incentive for preferring Wick's normal order within a bosonic context was to single out the energy contribution of the vacuum state (zero-point energy) to a unique term in the second-quantized Hamiltonian (the one involving the identity operator), we recall here that it also occurs in practice to prevent ill-defined situations due to an incomplete resolution of the identity for a truncated computational basis when assembling $\hat{H}_{M}$ algebraically as a sum of products of a limited number of elementary projected operators (or in matrix form, from their finite matrix representations).
While such a prescription is relevant in classical computing too, it may prove essential within a QC context if one wants to rely algorithmically-wise on the sole knowledge of the encoded action of $\hat{b}^{\dagger}_{M}$ with respect to the computational basis, as exemplified in the main text for unary or binary mappings.
Again, such a predicament is rarely encountered in application cases, simply because Wick's normal order has become customary.
Yet, one may be tempted to believe that the wrong choice has limited consequences, and we thus find it important to stress out that using the normal order should not be viewed as a merely convenient option.
It actually is critical and for reasons that are rarely discussed in the literature.

\section{Binary qubit-mapping of the bosonic Hamiltonian}

\subsection{Redundant Pauli terms}\label{app:Hqubit}

The so-called compact mapping~\cite{huang2022qubitization} associates the bosonic Fock states to qubit strings according to a numeral binary mapping.
For a total number of qubits denoted $ K $, and starting counting from right to left, we have the following mapping, 
\begin{eqnarray}
|0 \rangle  &=& | 0_{K}, 0_{K-1},...,0_{2}, 0_{1}   \rangle \quad, \nonumber\\
|1 \rangle  &=& | 0_{K}, 0_{K-1},...,0_{2}, 1_{1}   \rangle \quad, \nonumber\\
|2 \rangle  &=& | 0_{K}, 0_{K-1},...,1_{2}, 0_{1}   \rangle \quad, \nonumber\\
|3 \rangle  &=& | 0_{K}, 0_{K-1},...,1_{2}, 1_{1}   \rangle \quad, \nonumber\\
& \vdots & \nonumber\\
|2^K-1 \rangle  &=& | 1_{K}, 1_{K-1},...,1_{2}, 1_{1}   \rangle \quad.
 \label{eq:bitstring}
\end{eqnarray}

In order to show why the ladder operator in
Eqs.~(\ref{eq:op_ladder}-\ref{eq:binary_rep}) is decomposed into a linear combination of $O(M_l \log_2(M_l))$ (potentially nonlocal) Pauli strings only,
we have to count the unique types of bit-flip patterns in all the $(M_l-1)$ first-neighbor transition (jump) operators, $\ket{r_l+1}\bra{r_l}$.
We shall simply call them projectors in what follows.
The key insight is that the number of projectors having a unique type of bit-flip pattern corresponds to the number of possible lengths of trailing 1s within a $K$-bit binary integer, which is exactly $K = \log_2(M_l)$, \emph{i.e.}, following a logarithmic scaling with the number of basis functions.
Indeed, adding 1 to a binary number primarily affects the rightmost 0-bit by flipping it to 1. 
However, if there is a sequence of trailing 1s, all those 1s flip to 0, and the first 0-bit that is encountered flips to 1. 
Since a $K$-bit number can have at most $K$ different lengths of trailing 1s (ranging from 0 to $K-1$ trailing 1s), the number of unique transition patterns is at most $K = \log_2(M_l)$. 
Thus, there are exactly $\log_2(M_l)$ different patterns of bit flips, each of them giving $2^K = M_l$ different Pauli strings, finally totalling into $M_l\log_2(M_l)$ unique Pauli strings.
Let us give an example for three qubits. The number of projectors $\ket{ r_l + 1 }\bra{ r_l }$ is $2^3 - 1 = 7$.
They are given as 
\begin{eqnarray}
\lbrace &&\ket{001}\bra{000},
\ket{010}\bra{001},
\ket{011}\bra{010},
\ket{100}\bra{011}, \nonumber \\
&&
\ket{101}\bra{100},
\ket{110}\bra{101},
\ket{111}\bra{110}\rbrace \quad. 
\end{eqnarray}
Only three unique patterns of bit flips arise from these seven projectors, grouped as follows.
First, the first qubit passes from state 0 to 1, thus leading to, using Eq.~(\ref{eq:projectors_to_pauli}),
\begin{eqnarray}\label{eq:first_change}
&&\lbrace \ket{001}\bra{000}, \ket{011}\bra{010}, \ket{101}\bra{100}, \ket{111}\bra{110} \rbrace
\nonumber \\
&& \rightarrow (I \pm Z ) \otimes
(I \pm Z ) \otimes (X - iY) \quad.
\end{eqnarray}
Hence, those four different projectors share the same $2^3=8$ Pauli strings, up to relative phase factors within their linear combinations.
Second, the state of the first qubit passes from 1 to 0, and the second from 0 to 1, thus leading to
\begin{eqnarray}\label{eq:second_change}
&&\lbrace \ket{010}\bra{001}, \ket{110}\bra{101} \rbrace 
\nonumber \\
&&\rightarrow (I \pm Z ) \otimes (X - iY ) \otimes (X + iY) \quad,
\end{eqnarray}
\emph{i.e.}, two projectors share the same eight Pauli strings.
Finally, the last projector corresponds to a change in the state of the three qubits,
\begin{eqnarray}\label{eq:third_change}
\ket{100}\bra{011} = (X - iY ) \otimes (X + iY ) \otimes (X + iY) \quad,
\end{eqnarray}
and thus does not share any of its eight possible Pauli strings with another projector.
It is now clear from Eqs.~(\ref{eq:first_change}), (\ref{eq:second_change}), and (\ref{eq:third_change}) that the creation ladder operator can be written as a linear combination of at most $M_l \log_2(M_l)$ Pauli strings, which can be
recovered from the recursive relation in Eq.~(\ref{eq:Binary_bdag}). 
As a matter of example, the Hamiltonian models given in Eqs.~(\ref{eq:hunordered}-\ref{eq:hordered}) thus correspond to the two following Hamiltonian matrices with 35 terms,
\begin{eqnarray}
H_{\text{unordered}} &=& 
7355.49 \, III 
-1070.32\, IZZ 
-3226.57\, ZII \nonumber\\
&&-1046.88\, ZIZ 
-1054.69\, ZZI 
+ 1093.75\, ZZZ \nonumber\\
&&-1039.07 \,IZI 
-1536.13 \,IXX 
-787.44 \,IYY \nonumber\\
&&+ 1052.87\, ZXX
+ 520.69 \,ZYY 
- 350.32\, XIX \nonumber\\
&&+ 133.81\, XZX 
-350.32 \,YIY 
+ 133.81\, YZY \nonumber\\
&&-701.5\, XXX  
+ 359.17\, XYY 
-359.17\, YXY \nonumber\\
&&-701.5 \,YYX 
-1911.94 \,IIX
+ 297.03 \,IZX \nonumber\\
&&+ 1090.4 \,ZIX 
+ 259.3 \,ZZX + 
62.31 \,XII \nonumber\\
&&-15.69\, XIZ 
-31.35 \,XZI 
+ 3.86 \,XZZ \nonumber\\
&&+  263.80 \,IXI 
-11.88 \,IXZ 
-199.39 \,ZXI \nonumber\\ 
&&-19.39 \,ZXZ 
+ 125.97\, XXI 
-31.25 \,XXZ \nonumber\\
&&+  125.97\, YYI 
- 31.25\, YYZ \quad,
\end{eqnarray}
and
\begin{eqnarray}
H_{\text{ordered}} &=& 
8503.93 \,III 
- 1093.75\, IIZ
-2187.51\,IZI \nonumber \\ &&
-4375.01\, ZII 
-1536.13 \,IXX 
-787.43 \,IYY \nonumber \\ &&
+1052.87\, ZXX
+ 520.69 \,ZYY 
-350.32 \,XIX \nonumber \\ &&
+ 133.81\, XZX
-350.32 \,YIY
+ 133.81\, YZY \nonumber \\ &&
-701.5\,XXX 
+ 359.17\, XYY 
-359.17 \,YXY  \nonumber \\ &&
-701.5 \,YYX
-2379.65 \,IIX 
+ 764.74 \,IZX \nonumber \\ &&
+ 1558.14\, ZIX 
-208.4 \,ZZX 
+ 62.31 \,XII \nonumber \\ &&
-15.69 \,XIZ  
-31.35 \,XZI 
+ 3.86 \,XZZ \nonumber \\ &&
+ 314.43 \,IXI
- 62.51 \,IXZ 
-250.02\, ZXI \nonumber \\ &&
+  31.24\, ZXZ 
+ 125.97 \,XXI 
-31.25 \,XXZ \nonumber \\ &&
+  125.97 \,YYI  
-31.25 \,YYZ  
+ 23.44\, IZZ \nonumber \\ &&
+ 46.88\, ZIZ 
+ 93.75\, ZZI \quad,
\end{eqnarray}
which are obviously different, as expected from our previous analysis. 
In this, we used an energy scaling factor $\hbar\omega \equiv 2000$ $\text{cm}^{-1}$ and, for notational simplicity, we dropped out the Kronecker direct product symbol and omitted the qubit indices within the ordered strings.

\subsection{Origin of Eq.~(\ref{eq:Binary_bdag})}\label{app:origin_binary}

Let us start with $K = 2$ qubits,
which leads to Eq.~(\ref{eq:Ex_binary_bdag2}).
In this equation, the bosonic ladder operator
plays the role of a linear combination of  first-neighbor transitions of the form
\begin{eqnarray}
\hat{b}^\dagger_2
= \ket{1}\bra{0}
+ \sqrt{2}\, \ket{2}\bra{1}
+  \sqrt{3}\, \ket{3}\bra{2} \quad ,
\end{eqnarray}
which reads within the binary encoding,
\begin{eqnarray}
\hat{b}^\dagger_2
= \ket{01}\bra{00}
+ \sqrt{2}\, \ket{10}\bra{01}
+  \sqrt{3}\, \ket{11}\bra{10} \quad ,
\end{eqnarray}
where each projector is associated to an operator $\hat{c}_i^{(1,2)}$.
Now, if we increase the number of qubits by one, it means that we add a qubit on the left side of $\ket{\eta_2,\eta_1}$, \emph{i.e.}, $\ket{\eta_3,\eta_2,\eta_1} \equiv \ket{\eta_3}\otimes \ket{\eta_2,\eta_1} $, where
$\eta_i \in \lbrace 0, 1 \rbrace$.
This additional qubit doubles the number of projectors, plus one, passing from 3 to 7 in this case.
These projectors can be divided into 3 groups: (1)
a group of $K-1$ projectors where the additional qubit is in state 0 and nothing changes for the projector of the previous qubits,
(2) a group of a single projector where the additional qubit is in state 1 and every other qubits are passing from state 1 to state 0,
(3) a group of $K-1$ projectors where the additional qubit is in state 1 and nothing changes for the projector of the previous qubits,
\begin{eqnarray}
\hat{b}^\dagger_3
= \ket{0}\bra{0} &\otimes& \left(\ket{01}\bra{00}
+ \sqrt{2}\, \ket{10}\bra{01}
+  \sqrt{3}\, \ket{11}\bra{10}\right) \nonumber \\
+ \ket{1}\bra{0} &\otimes& \sqrt{4}  \, \ket{00}\bra{11} \nonumber\\
+ \ket{1}\bra{1} &\otimes& \left(
\sqrt{5} \ket{01}\bra{00}
+ \sqrt{6}\, \ket{10}\bra{01}
+  \sqrt{7}\, \ket{11}\bra{10}\right). \nonumber \\
\end{eqnarray}
These are the only possible first-neighbor transitions.
This can be generalised easily to $K$ qubits
where,
according to Eq.~(\ref{eq:projectors_to_pauli}),
(1) the first group implies
applying $I^+$ on the $K$-th qubit
while keeping the same operators $\hat{c}_i^{(1, K-1)}$ on the $(K-1)$ ones,
(2) the second group
leads to $\sigma^+$ on the $(K-1)$ previous qubits and an additional $\sigma^-$ on the $K$-th qubit,
and
(3) the third group implies
applying $I^-$ on the $K$-th qubit
while keeping the same operators $\hat{c}_i^{(1, K-1)}$ on the $(K-1)$ ones.
One can now readily associate each group to the final formula in Eq.~(\ref{eq:Binary_bdag}).



\newcommand{\Aa}[0]{Aa}

\end{document}